\documentclass[Style/sn-mathphys]{Style/sn-jnl}[8pt]
\UseRawInputEncoding
\graphicspath{{Figures/}}
\usepackage{amsmath}

\usepackage{etoolbox,xspace}
\usepackage{booktabs}
\usepackage{physics}

\newcommand{\definetextttcmd}[1]{\csdef{#1}{\texttt{#1}\xspace}}
\forcsvlist{\definetextttcmd}{PORTALS, CGYRO, GYRO, TGLF, SAT2, TGYRO, NEO, BoTorch, GPyTorch, ASTRA, TRANSP, GX, TANGO, GENE, TRINITY, QuaLiKiz, PyTorch, GKW, NCLASS, EPED, TORIC, FPPMOD, MAESTRO, FreeGS, TEQ, ALCON, quends, NUBEAM}

\begin{document}

\title[Impact of energetic alpha particles on core turbulence]{Impact of energetic alpha particles on core turbulence in an ARC-class fusion power plant}

\author*[1]{\fnm{J.} \sur{Hall}}\email{hallefkt@mit.edu}
\author[1]{\fnm{N. T.} \sur{Howard}}
\author[1]{\fnm{P.} \sur{Rodriguez-Fernandez}}
%\author[1]{\fnm{L.} \sur{Nichols}}
\author[1]{\fnm{R.A.} \sur{Tinguely}}
\author[2]{\fnm{I.} \sur{Sfiligoi}}
\author[3,4]{\fnm{J.} \sur{Ruiz-Ruiz}}
\author[5]{\fnm{J.C.} \sur{Hillesheim}}
\author[5]{\fnm{A.} \sur{Creely}}
\author[6]{\fnm{E.A.} \sur{Belli}}
\author[6]{\fnm{J.} \sur{Candy}}

\affil[1]{\orgdiv{MIT Plasma Science and Fusion Center}, \city{Cambridge}, \state{MA}, \country{USA}}
\affil[2]{\orgdiv{University of San Diego}, \city{La Jolla}, \state{CA}, \country{USA}}
\affil[3]{\orgdiv{York Plasma Institute}, \city{Heslington}, \country{UK}}
\affil[4]{\orgdiv{Rudolf Peierls Centre for Theoretical Physics}, \city{Oxford}, \country{UK}}
\affil[5]{\orgdiv{Commonwealth Fusion Systems}, \city{Devens}, \state{MA}, \country{USA}}
\affil[6]{\orgdiv{General Atomics}, \city{San Diego}, \state{CA}, \country{USA}}

\abstract{In this work, we investigate the impact of fusion-born alpha particles on core turbulence and transport in the ARC tokamak fusion power plant \cite{hillesheimOverviewPhysicsBasis2026} using linear and nonlinear gyrokinetic \CGYRO\cite{candyHighaccuracyEulerianGyrokinetic2016} simulations. A significant reduction in ion-scale turbulent heat and particle fluxes is observed in the inner core (r/a $\leq$ 0.5), which is associated with multiscale interactions between fast ion-destabilized modes, zonal flows, and the background turbulence. A nonlinear upshift in the ITG critical gradient is observed in the simulations with fast alphas compared to those with artificially thermalized alphas. The turbulence reduction is found to scale beneficially with alpha particle density and plasma $\beta_e$, and the radial extent of the turbulence suppression is limited to the volume containing a significant density of fast particles. The suitability of local gyrokinetics and potential impacts of fast ion effects on fusion performance are discussed.}

\keywords{
Burning plasma, ARC, alpha particles, gyrokinetics, CGYRO, turbulence stabilization,  fast ions, TAEs, zonal flows.
\\ 
\\
\textit{Submitted for publication.}
}

\maketitle
\section{Introduction}
\label{sec:Introduction}

Burning plasma conditions in tokamaks will be characterized by a large population of fusion-born alpha particles. It is not yet well-understood how these fast ions will affect main ion heat and particle transport, and therefore overall performance. Recent advances across theory, computation, and experiment suggest that tokamak plasmas with a large fast ion population may exhibit reduced core turbulence and increased performance \cite{citrinOverviewTokamakTurbulence2023,naHowFastIons2025}. This is vital to understand for burning plasma devices, in breakeven-class ($Q\geq1$) devices such as SPARC \cite{creelyOverviewSPARCTokamak2020} and ITER \cite{barabaschiProgressITERIts2026}, but even more so in proposed fusion power plants (FPP) such as ARC \cite{hillesheimOverviewPhysicsBasis2026}, where alpha particles will be the dominant source of heating. Recent work has predicted the fusion performance of the ARC ``V3A" design across several levels of model fidelity, including direct profile prediction with nonlinear gyrokinetics \cite{howardPerformanceTransportARC2026}, building on the development of direct \cite{candyTokamakProfilePrediction2009} and surrogate-based \cite{rodriguez-fernandezNonlinearGyrokineticPredictions2022} gyrokinetic profile prediction methods. Here we address one of the key physics gaps in local gyrokinetic modeling of this ARC design, namely the effect of including a fast alpha population in nonlinear simulations.

The possibility that fast particles could experience anomalous transport was recognized early by Rosenbluth and Rutherford \cite{rosenbluthExcitationAlfvenWaves1975}, who outlined the mechanism by which high-energy beam ions can destabilize shear Alfven waves in a tokamak. For many years the dominant concern was that alpha-driven Alfvenic instabilities, particularly Toroidal Alfven Eigenmodes (TAE) \cite{fuExcitationToroidicityinducedShear1989, CandyJ1993AntA,rosenbluthContinuumDampingHighmodenumber1992, zoncaResonantDampingToroidicityinduced1992} would degrade overall reactor confinement. In the next decade, nonlinear gyrokinetic results \cite{estrada-milaTurbulentTransportAlpha2006} demonstrated that alphas will also be transported by core turbulence. It is now understood that fast ions may also \textsl{improve} confinement through various mechanisms. For example, fast ions increase the normalized pressure gradient $\beta^*$ and Shafranov shift, which suppresses drift-wave microturbulence \cite{bourdelleStabilizingImpactHigh2003, belliFullyElectromagneticGyrokinetic2010,garciaKeyImpactFinitebeta2015} and has been connected to fast-ion and alpha-particle pressure stabilization in reactor-relevant hybrid scenarios \cite{romanelliFastIonStabilization2010}. If the fast ion population is large enough to impact the dilution of the background plasma, this can reduce the turbulence drive and increase performance, even with a lower main ion fraction, as has been experimentally demonstrated in KSTAR \cite{kimTurbulenceStabilizationTokamak2023,naFIREModeKSTAR2026}. Another stabilizing mechanism is a resonant interaction \cite{sienaFastionStabilizationTokamak2018, disienaResonantInteractionEnergetic2019} occurring when the unstable mode frequencies are comparable to the fast ion drift frequency $\omega_{d\alpha}$.

In burning plasmas with MeV alpha particles, these ``direct" mechanisms seem less likely to play a major role. The $\beta^*$ stabilization is minor because the fast ion pressure represents only a small fraction of the total at the high temperatures and densities of an FPP; we demonstrate in this paper that it causes only minor reductions in turbulent heat and particle fluxes in ARC. The turbulence stabilization effect of high dilution will also likely not play a significant role, as diluting the fuel ions is counter to high fusion performance (except possibly for certain near breakeven-conditions, as was demonstrated in \cite{rodriguez-fernandezCorePerformancePredictions2024}). The resonant interaction is suppressed because the fast-ion (EP) and thermal (ITG) modes occupy well-separated wavenumber ranges ($k_y\rho_s \gg k_y\rho_\alpha$) such that distinct ITG and EP modes coexist. The resonant mechanism can dominate in the complementary regime of steep fast ion temperature gradient and shallow density gradient \cite{wilkieFirstPrinciplesModelling2018, disienaPredictionsImprovedConfinement2023}, typical of ICRH-accelerated minority ions but not of fusion-born alphas with their strong density gradients.

Finally there is zonal-flow enhancement, whereby fast ion-driven instabilities are excited at long wavelength, $k_y\rho_\alpha \ll k_y \rho_s$ due to the large fast ion gyroradius $\rho_\alpha \gg \rho_s$, and consequently couple to zonal flows with much greater efficiency than ion-scale turbulence, driving enhanced zonal flow shear that suppresses heat transport \cite{disienaElectromagneticTurbulenceSuppression2019, garciaStableDeuteriumTritiumPlasmas2024}. These ``indirect" fast ion stabilization effects are expected to be the dominant mechanism by which fast particles interact with the turbulence in future devices \cite{naHowFastIons2025}. Fusion alphas are particularly relevant due to the alpha particle birth energy having a thermal velocity $v_{\alpha,0}>v_A$ in most devices, including ARC, where $v_A$ the on-axis Alfven speed. As these alphas slow down primarily via collisions they can destabilize Alfvenic modes as they cross the condition $v_{\alpha}/v_A$. TAE modes in particular have long been known to nonlinearly drive a net zonal flow through analytic theory \cite{chenNonlinearExcitationsZonal2012, qiuEffectsEnergeticParticles2016,qiuNonlinearExcitationFiniteradialscale2017} and through simulation \cite{zhihonglinNonlinearGenerationZonal2013,chenEffectsZonalFields2024,garciaStableDeuteriumTritiumPlasmas2024}, which has been confirmed experimentally in both JET \cite{ruizruizMeasurementZeroFrequencyFluctuations2025} and DIII-D \cite{duFirstMeasurementDriftAlfven2024}.

Zonal-flow enhancement and turbulence reduction by fast ion-driven modes has been observed experimentally in multiple devices \cite{mazziEnhancedPerformanceFusion2022, duAlfvenEigenmodesSuppress2025, disienaAssessingImpactAlpha2024}. However, there has been comparatively little work dedicated to predictive analysis of burning plasma FPP designs using high-fidelity nonlinear gyrokinetic simulations, which are required to capture the correct quantitative thermal and fast-ion dynamics. Studies performed in burning plasma conditions have mainly focused on next-step devices with $Q\leq10$, $P_\alpha\sim P_{\text{aux}}$ \cite{sienaPredictionsImprovedConfinement2023,tangGyrokineticSimulationEffects2025}, as opposed to an FPP with $P_\alpha\gg P_\text{aux}$. In this work, we examine the impacts of fast alphas on core turbulence and transport in a burning plasma ($P_{\text{fus}}=500$ MW, $Q=20$) ARC scenario. We refer to the ``V3A'' engineering design presented in \cite{hillesheimOverviewPhysicsBasis2026}.

The remainder of this paper proceeds as follows: in Section \ref{sec:ARC Medium fidelity case}, we describe the reduced-current ARC scenario under investigation. In Section \ref{sec:Linear CGYRO Modeling}, linear \CGYRO simulations are reported for various radial locations both with and without fast alphas, and the mode structure and polarization of the observed fast ion instabilities are discussed. In Section \ref{sec:Nonlinear CGYRO Modeling}, we detail nonlinear \CGYRO simulations across different values of radial location, normalized gradients, alpha particle density and density gradient, and $\beta_e$. We conclude by discussing the potential impact of the observed turbulence stabilization on device performance.

\section{ARC Medium fidelity case}
\label{sec:ARC Medium fidelity case}

In contrast to breakeven-class experimental devices like SPARC \cite{creelyOverviewSPARCTokamak2020} and ITER \cite{barabaschiProgressITERIts2026}, a commercially viable FPP will need to attain even higher values of fusion gain $Q$ and alpha heating power $P_\alpha$. The successful design of an FPP will thus depend largely on effective physics-based predictions of core plasma turbulence and transport, since the conditions in a burning plasma are not currently accessible to experimental devices. ARC \cite{hillesheimOverviewPhysicsBasis2026} is a proposed tokamak power plant concept being designed by Commonwealth Fusion Systems in collaboration with public research for construction in the early 2030s. Recent work has been published on the tokamak physics basis for ARC, including the predictions for fusion performance with a wide range of low-, medium- and high-fidelity modeling tools \cite{howardPerformanceTransportARC2026}. Recent advances in integrated modeling, optimization, and profile predictions with nonlinear gyrokinetics has led to ARC being one of the first fusion power plant concepts to be designed from the beginning using first-principles transport physics. The ARC baseline operational point is a 400 MWe ($\sim$1 GWth), major radius $R_0=4.62$ m, minor radius $a=1.18$ m, fully inductive (plasma current $I_p=12$ MA), high-field ($B_0=11.4$ T) tokamak employing high temperature superconducting magnets, a liquid immersion molten salt blanket for power conversion and tritium self-sufficiency, and up to $50$ MW of Hydrogen minority ICRH heating power. ARC will be a strongly burning plasma, with a plasma gain of $Q\sim50$, making it vital to understand the potential impacts of alpha particles and fast-ion instabilities on the overall device performance.

In this work, we investigate a modified ARC scenario relative the V3A design point outlined in \cite{hillesheimOverviewPhysicsBasis2026} and described above. Due to the high computational cost of nonlinear gyrokinetic simulations, and in particular the difficulty of local gyrokinetic simulations with low magnetic shear, which is a feature of the ARC equilibrium due to the long pulse time and lack of external current drive, we present an initial analysis here of a reduced-current ($9$ MA) ARC scenario. In addition to the lower computational cost, a reduced-current scenario is favorable for this type of analysis because it gives a helpful indication of whether significant alpha-particle effects on turbulence are visible at lower fusion gain factors of $Q\sim20$, and whether we expect such effects to scale favorably with overall fusion power.

\begin{table}
\begin{tabular}{@{}lllll@{}}
\toprule
\textbf{Input Parameters} &  &  & \textbf{Outputs} &  \\
\midrule
$R$ & $4.62$ m &  & $P_{fus}$ & $511$ MW \\
$a$ & $1.18$ m &  & $H_{98}$  & $1.05$ \\
$I_p$ & $9$ MA &  & Q & $21.7$ \\
$B_t$ & $11.4$ T &  & $\langle T_e\rangle$ & $8.0$ keV  \\
$\kappa_{\text{sep}}$ & $1.89$ &  & $\langle n_e\rangle$ & $2.3 \times 10^{20}$ m$^{-3}$ \\
$\delta_{\text{sep}}$ & $0.56$ &  & $f_{G,\text{ped}}$ & $1.01$ \\
$P_{\text{aux}}$ & $23.5$ MW &  & $q_{95}$ & $5.07$ \\
$Z_{\text{eff}}$ & $1.5$ &  &  &  \\ 
\bottomrule
\end{tabular}

\caption{Engineering parameters and plasma assumptions used for \MAESTRO medium fidelity integrated modeling, and the resulting fusion performance prediction.}
\label{tab:table1}
\end{table}

\begin{figure*}
    \centering
    \includegraphics[width=\linewidth]{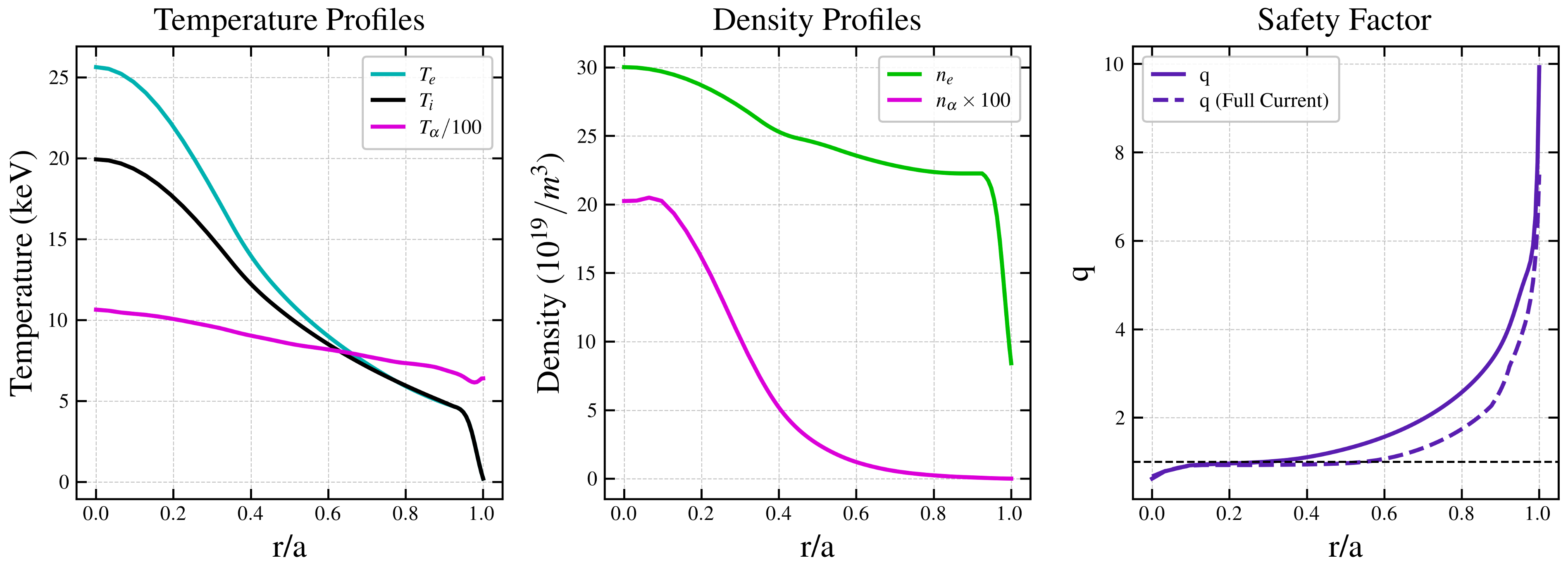}
    \caption{Kinetic profiles predicted through \MAESTRO medium fidelity integrated modeling. Scaled alpha particle temperature and density profiles predicted with \NUBEAM are plotted in magenta. The $q$-profile from the full-current discharge reported in \cite{hillesheimOverviewPhysicsBasis2026} is shown for reference.}
    \label{fig:medfid}
\end{figure*}

In order to provide input profiles for linear and nonlinear gyrokinetic analysis with \CGYRO, a medium-fidelity performance prediction was performed with the \MAESTRO integrated modeling tool \cite{rodriguez-fernandezAcceleratingIntegratedModeling2026a}. In this context, ``medium fidelity" refers to predictions carried out with physics-based transport models, such as quasilinear models like TGLF \cite{staeblerTheorybasedTransportModel2007} and Qualikiz \cite{bourdelleCoreTurbulentTransport2015}, which rely on saturation rules or free parameters determined from nonlinear gyrokinetics. In contrast, ``high fidelity" simulations in this context refer to transport predictions directly utilizing nonlinear gyrokinetics. A neural network surrogate model of \EPED \cite{snyderFirstprinciplesPredictiveModel2011, muracaIntegratedModelingSPARC2025,hallMultiFidelityPredictiveCore2024} was used to self-consistently update the temperature boundary condition at pedestal top at each step in the transport simulation, assuming a fixed electron density of $2.1\times10^{20}$ at the pedestal top and an enforced ratio of separatrix electron density to pedestal electron density of $0.4$, both of which were chosen for favorable access conditions to peeling-limited small-ELM regimes \cite{eichPowerParticleExhaust2026}. From the pedestal top inwards, the \PORTALS transport solver \cite{rodriguez-fernandezEnhancingPredictiveCapabilities2024a} was used to predict the steady-state temperature, density, and current profiles using the \TGLF-\texttt{SAT2} \cite{staeblerGeometryDependenceFluctuation2021} quasilinear model for turbulent transport and \NEO \cite{belliKineticCalculationNeoclassical2008} for neoclassical transport. The thermal species simulated included electrons, Deuterium, Tritium (with $n_D=n_T$), a lumped low-Z impurity (Z=5), and Tungsten (Z=48). Interpretive \TRANSP \cite{pankinTRANSPIntegratedModeling2025} simulations were used to predict equilibrium, sawtooth crashes \cite{porcelliModelSawtoothPeriod1996}, ICRH heating and fast ion profiles. Full-wave \TORIC-\FPPMOD simulations \cite{brambillaNumericalSimulationIon1999, hammettFASTIONSTUDIES} were used to model the ICRH power deposition. The alpha particle effective temperature and density profiles were calculated using high-resolution \NUBEAM simulations \cite{pankinTokamakMonteCarlo2004} ($10^6$ MC particles) using the standard collisional slowing-down model. For the purposes of gyrokinetic simulations, the distribution function was assumed to be an effective Maxwellian, similar to in recent work also studying alpha particles in burning plasma conditions \cite{disienaFirstGlobalGyrokinetic2025}. Whether assuming an effective Maxwellian distribution for the alphas changes the stability of TAEs and other modes, or affects global performance through alpha redistribution, is an important problem left to future studies.
For the engineering parameters and input assumptions in Table \ref{tab:table1}, the predicted kinetic and fast ion temperature and density profiles are presented in Figure \ref{fig:medfid}. At 75\% $I_p$ relative to the full-current ARC V3A design, \MAESTRO predicts $P_{fus}=511$MW ($Q$=21.7), compared to $\sim900$MW ($Q$=38.9) for the full current scenario \cite{howardPerformanceTransportARC2026}. A moderately sheared, monotonic $q$-profile ($q_{95}=5.07$) was obtained after allowing \TRANSP to evolve current diffusion over 20 s of plasma time. Longer current diffusion times did not affect $q$-profile shape or fusion performance with this modeling framework. Because of the reduced current, a pedestal top Greenwald density fraction \cite{greenwaldDensityLimitsToroidal2002} of $f_G=1.01$ was obtained. Recent work on high power metal-wall tokamaks additionally suggests that the classical Greenwald density limit may be conservative estimate with high heat flux at the separatrix \cite{giacominFirstPrinciplesDensityLimit2022}. However, for the present work this is merely an artifact of having changed only the plasma current while leaving all other operational parameters fixed for this specific case, and is not meant to suggest that operation near or above the Greenwald density limit is desired or expected in ARC.

\begin{figure}
    \centering
    \includegraphics[width=0.45\linewidth]{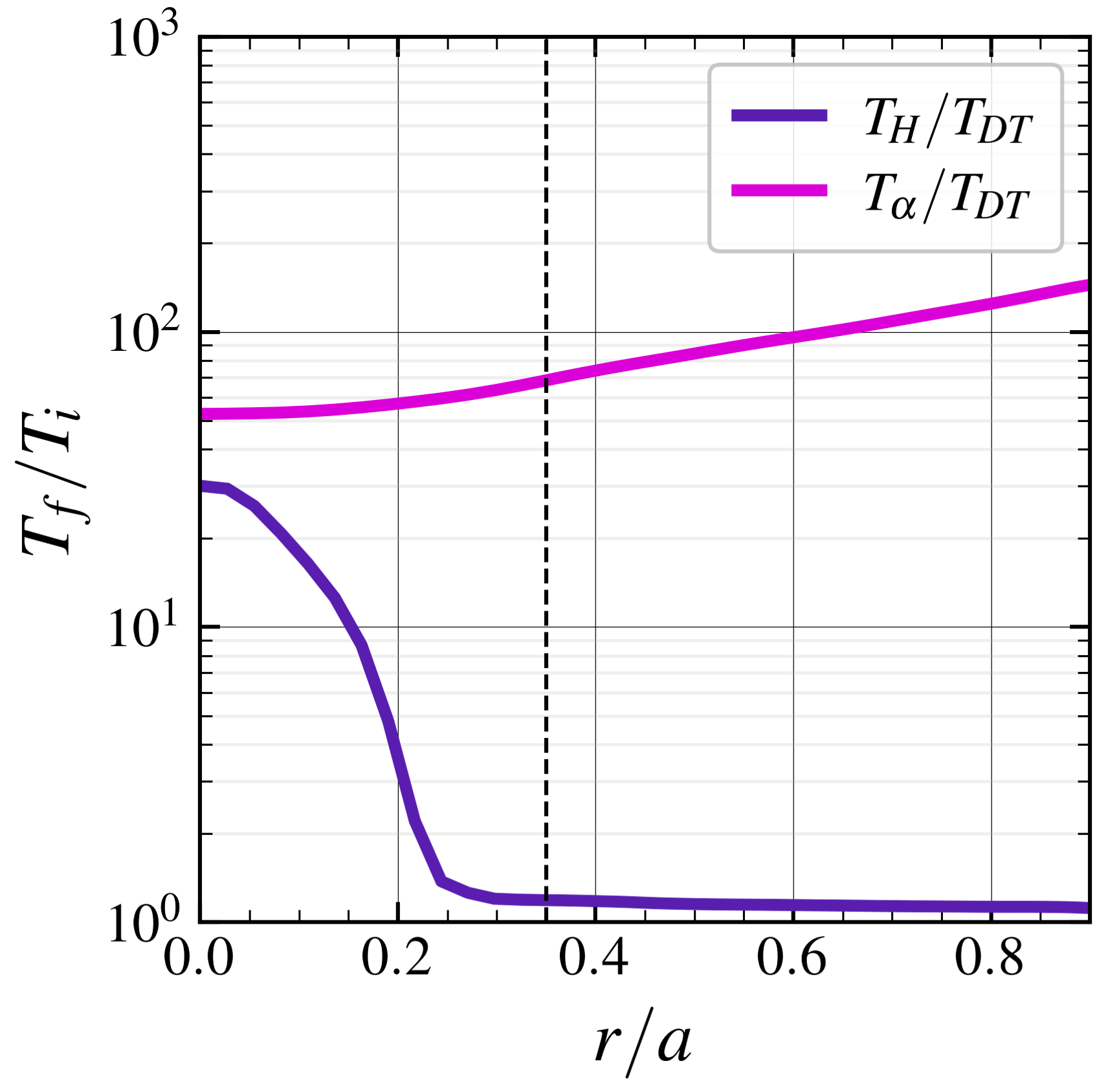}
    \caption{Ratio of fast ion to thermal (lumped DT) ion temperatures for ICRH-accelerated Hydrogen minority species predicted with \TORIC-\FPPMOD and fusion alphas predicted with \NUBEAM. In both cases the effective Maxwellian temperature output by \TRANSP is used. The treatment of fast Hydrogen as a thermal impurity is justified since $T_{H}\sim T_i\lt T_\alpha$ at all radial locations simulated with nonlinear \CGYRO.}
    \label{fig:fasttempratios}
\end{figure}

The \MAESTRO medium fidelity predictions include profiles of six thermal and fast species, D, T, $^4\text{He}$, $^1\text{H}$, W, and a lumped impurity (Z=5, A=10). In the linear and nonlinear \CGYRO simulations in the following sections, the Deuterium and Tritium species were lumped into one main ion species (Z=1, A=2.5, $f_{DT}=0.85$). Treating D and T as a lumped thermal species has been shown to be a reasonable approximation which significantly reduces computational cost. \cite{disienaAssessingImpactAlpha2024}. The Tungsten, minority Hydrogen and low-Z impurity were lumped into one species (Z=4, $f_{\text{imp}}=0.033$) by enforcing quasineutrality for $Z_{L}n_{L}=\sum Z_in_i$ subject to the constraint that $Z_{\text{eff}}=\sum Z_i^2n_i/n_e=\text{const}$. The lumped impurity charge was then rounded to the nearest integer $Z$, satisfying quasineutrality but causing a small ($\lt1\%$ for these profiles) change in $Z_{\text{eff}}$. The alpha particle effective temperature and density profiles are kept as the effective Maxwellian output by \TRANSP. The fast Hydrogen ions resulting from ICRH heating are artificially set equal to the thermal ion temperature, as their radial extent is highly localized to the magnetic axis as demonstrated in Figure \ref{fig:fasttempratios}. Neglecting the nearly-thermal (outside of $r/a\approx0.2$) fast Hydrogen and treating it instead as a trace impurity is likely to to produce a negligible effect on the Alfvenic stability as well as not inducing a significant change in total pressure at the radial locations of interest, as $p_\alpha\gg p_{H}$ is well satisfied. The main ion dilution of this scenario is low ($f_{DT}=0.85$), and the fast particle fraction ($f_{\alpha}\sim 10^{-3}$) in particular produces a negligible contribution to the overall dilution. It should be mentioned that Helium ash is not included in the lumped low-Z impurity, as a full treatment of ash thermalization and transport lies beyond the scope of this analysis. Throughout this work, we will refer to ``Control" and ``Fast Ion" simulations. In the ``control" simulations, the fast alpha species is artificially set equal to the (DT) main ion temperature. This ensures that any differences in turbulent heat fluxes observed are independent of effects due to dilution and $Z_{\text{eff}}$. Additionally, the 2D equilibrium is kept constant between the two cases despite the small change in pressure profiles.

\section{Linear \CGYRO Modeling}
\label{sec:Linear CGYRO Modeling}

Here, we report the results of linear \CGYRO modeling using the ARC profiles generated in the previous section. \CGYRO is a modern local, $\delta f$ Eulerian gyrokinetic code \cite{candyHighaccuracyEulerianGyrokinetic2016} which has been heavily optimized for modern GPU architectures \cite{candyMultiscaleoptimizedPlasmaTurbulence2019}. \CGYRO has been extensively validated against experiment \cite{whiteValidationNonlinearGyrokinetic2019, odstrcilDependenceImpurityTransport2020, howardMultiscaleGyrokineticSimulations2016} and has been used to successfully reproduce flux-matched experimental profiles, beginning with early work by \cite{candyTokamakProfilePrediction2009} using a traditional Newton solver approach, and more recently using surrogate-based optimization methods on DIII-D \cite{howardSimultaneousReproductionExperimental2024a} and ASDEX Upgrade \cite{bielajewGyrokineticProfilePrediction2025} using \PORTALS \cite{rodriguez-fernandezEnhancingPredictiveCapabilities2024}. 

\begin{figure*}
\centering
    \includegraphics[width=\linewidth]{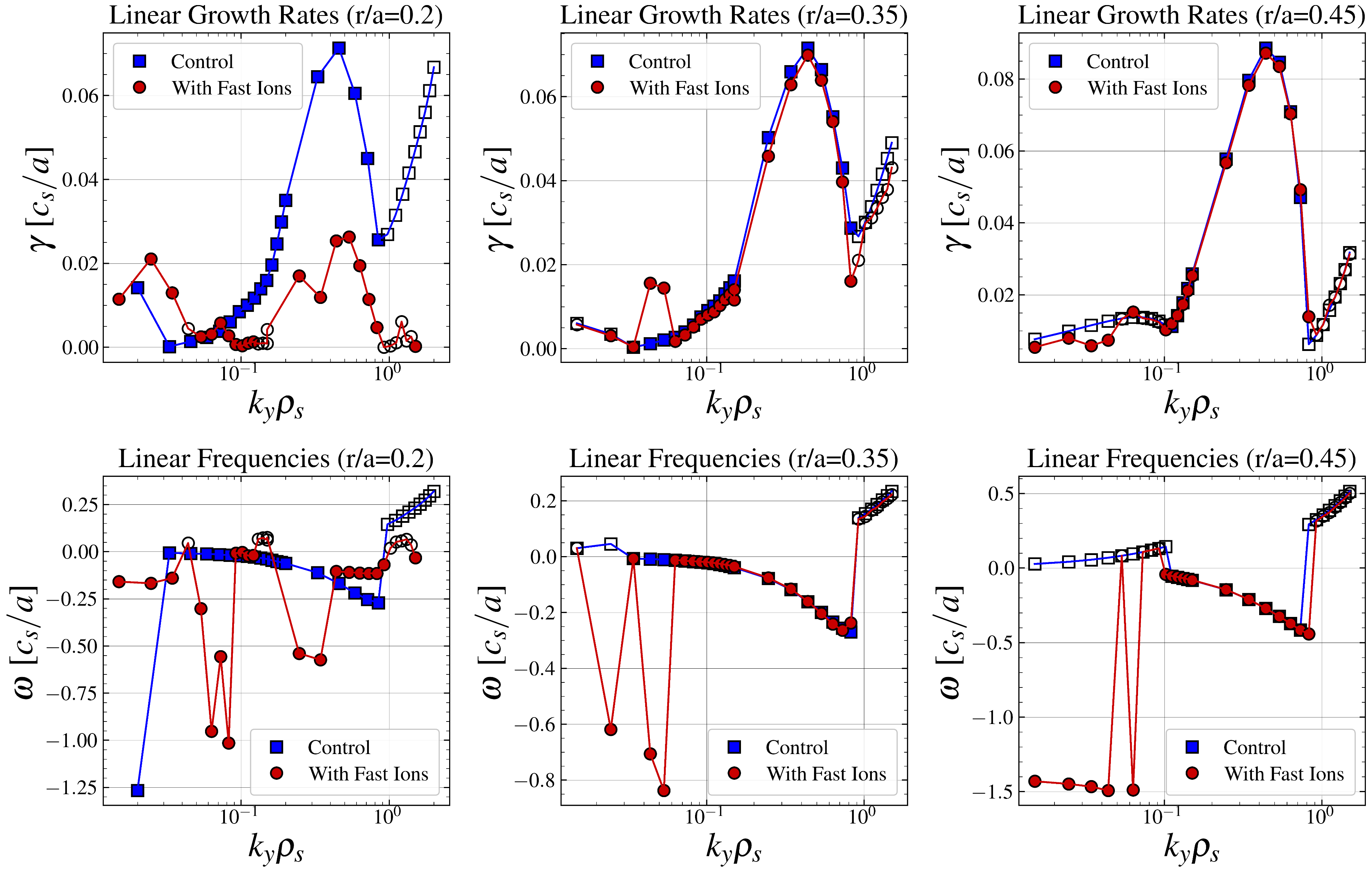}
    \caption{Linear growth rates calculated with \CGYRO with fast alpha profiles from \NUBEAM (``With Fast Ions") and with alphas set equal to the background ion temperature (``No Fast Ions"). Filled circles denote ion-diamagnetic directed modes (negative frequency), open circles denote electron-directed modes (positive frequency). At $r/a=0.2$, significant linear stabilization is observed in the ITG range of wavenumbers ($k_y\rho_s=[0.1,1]$). At $r/a=0.35$, very little linear effect is observed, but a peak in growth rates is seen at $k_y\rho_s\lt0.1$. At $r/a=0.45$, the alpha density is so low that there is negligible impact from fast particles.}
    \label{fig:linear}
\end{figure*}

Linear \CGYRO simulations were performed at three inner radial locations $r/a = [0.2,0.35,0.45]$, where $r=(R_+-R_-)/2$
and $a$ is the tokamak minor radius with $R_+$ and $R_-$ defined as the intersection of the flux surface with the midplane as in \cite{candyHighaccuracyEulerianGyrokinetic2016}. Linear simulations were carried out across a range in binormal wavenumbers $k_y\rho_s=[0.015,1.5]$, with radial, pitch angle, poloidal, and energy resolutions $N_r,N_\xi,N_\theta, N_\varepsilon = [64,16,24,8]$. All simulations were run until converged to a tolerance of $10^{-3}$, which required a large range of simulation times depending on the wave number and presence of fast ions. The low-$k_y$ locations with fast ions in particular required special attention due to the slow-growing fast-ion modes described in the following sections, typically requiring simulation times on the order of $10^4$ a/$c_s$ in order to reach the desired tolerances. For several wavenumbers and radial locations, it was necessary to increase the number of radial points to 96 to ensure that the electrostatic perturbation properly decayed to zero at the domain boundary. The resulting linear growth rates are plotted in Fig. \ref{fig:linear}. 

One key result from linear simulations is that the presence of a high fast particle population at $r/a=0.2$ has a strongly stabilizing effect in ion-scale growth rates. This linear effect is likely due in part to the significant alpha particle contribution to the the normalized pressure gradient $\beta^*=-8\pi n_eT_e/B^2_{unit}\frac{dp}{dr}$, where $B_{unit}=\frac{q}{r}\psi '$  at $r/a=0.2$, which is on the order of 25\%. In addition to the ion diamagnetic-directed modes, the electron diamagnetic-directed modes above $k_y \rho_s\sim1$ are almost completely stabilized with the inclusion of fast alpha particles. While these direct effects are important to consider, we chose not to investigate this radial location with nonlinear simulations. Physically, locations this close to the magnetic axis are often not well described solely by turbulence modeling, as the transport is often affected by MHD modes and sawteeth (this location is inside of the $q=1$ surface predicted with \TRANSP, as shown in Figure \ref{fig:medfid}), as well as practically being very difficult to saturate due to the small gradients and large simulation domains needed to properly resolve large-scale modes. Additionally, at this location, excluding the fast minority heating ions is likely to impact the results significantly, unlike for $r/a\geq0.3$ where we have already argued that treating it as a thermal species is appropriate. Finally, the plasma volume enclosed at this location is such a small contribution to the total that even if fast ions were shown to dramatically increase the gradients, it would not have a large impact on overall performance. 

\begin{figure}
    \centering
    \includegraphics[width=0.5\linewidth]{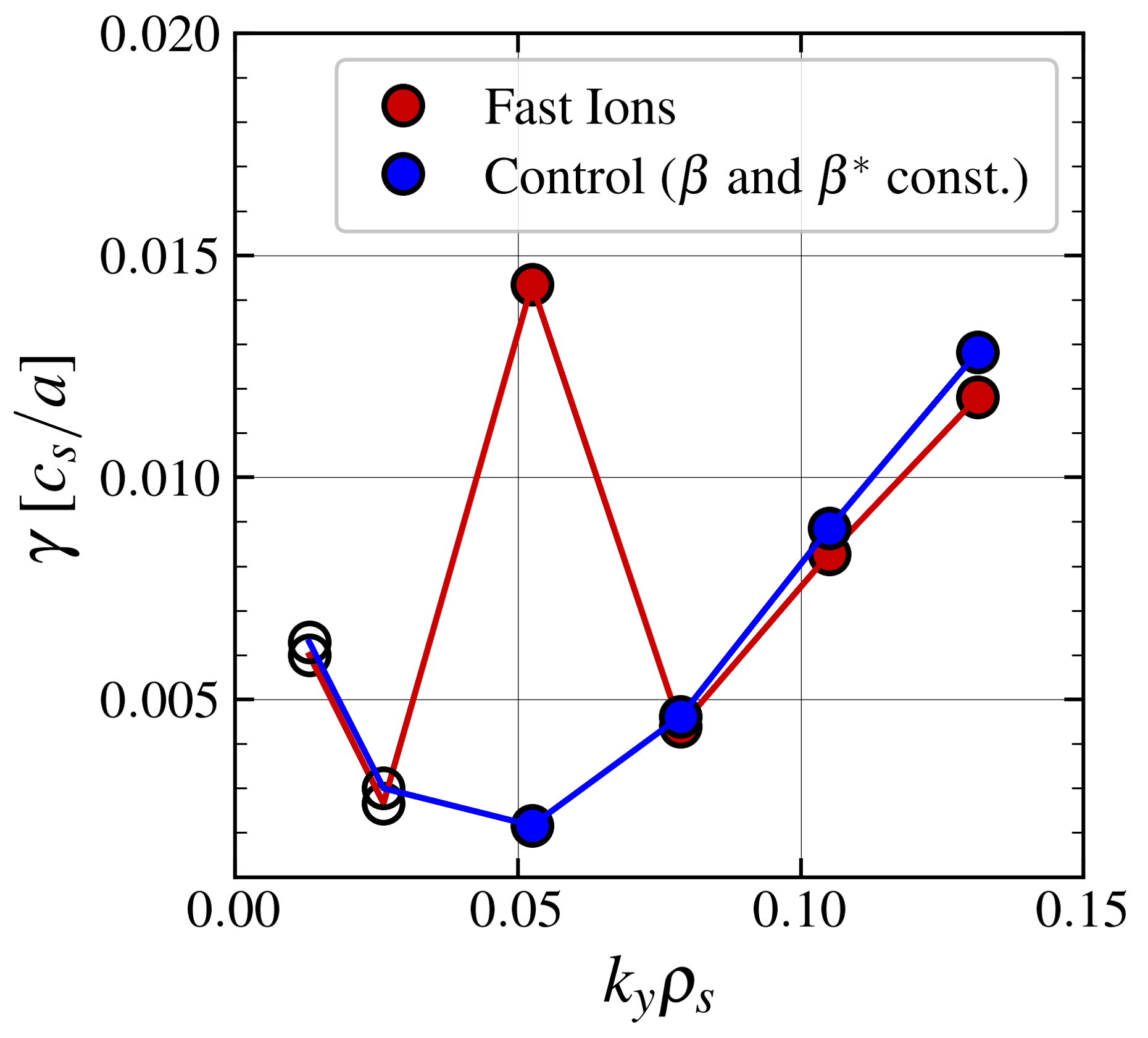}
    \caption{Linear scan in $k_y\rho_s$ at $r/a=0.35$. Even when the $\beta$ and normalized pressure gradient $\beta^*$ of the control simulation are increased to match the contribution from fast alphas, a decrease in the ITG growth rates for $k_y\rho_s>0.1$ is observed. Similarly, the fast ion mode at $k_y\rho_s>0.1$ is excited only when fast alphas are included independent of the background pressure profile.}
    \label{fig:constbeta}
\end{figure}

Linear simulations at $r/a=0.35$, by contrast, do not show a significant linear stabilization effect on ion or electron-directed modes, despite again modest contribution to the normalized pressure gradient by alphas of approximately 18\%. This is likely because the alpha density decreases significantly between $r/a=0.2$ and $r/a=0.35$. A direct comparison between the fast ion case and a case with thermal alphas with $\beta_e$ and $\beta^*$ artificially increased to match the fast alpha contribution shows that the change in growth rates can be seen only through changing the fast ion temperature, and that increased electromagnetic stabilization is not the dominant effect (Figure \ref{fig:constbeta}). An interesting feature at this radial location is the presence of fast alpha-destabilized modes at $k_y\rho_s\sim0.05$, corresponding to toroidal node number $n = 15-17$, calculated according to $n=k_{s}\rho_s / k_{s,0}\rho_s$, with $k_{y,0}\rho_s=q/r\rho_s$. The mode at $k_y\rho_s=0.052$ in particular will be discussed in the following sections. 

Finally, linear simulations conducted at $r/a=0.45$ show next to no impact on linear growth rates with the inclusion of fast particles, which is unsurprising; at this location, there is another factor of two reduction in the alpha particle density, corresponding to a fast alpha fraction of less than $10^{-3}$. Any differences between the ion scale growth rates between the two cases are close to numerical tolerances and considered insignificant. High-frequency fast ion-driven modes are observed, however their linear growth rates remain small.

\begin{figure}
    \centering
    \includegraphics[width=\linewidth]{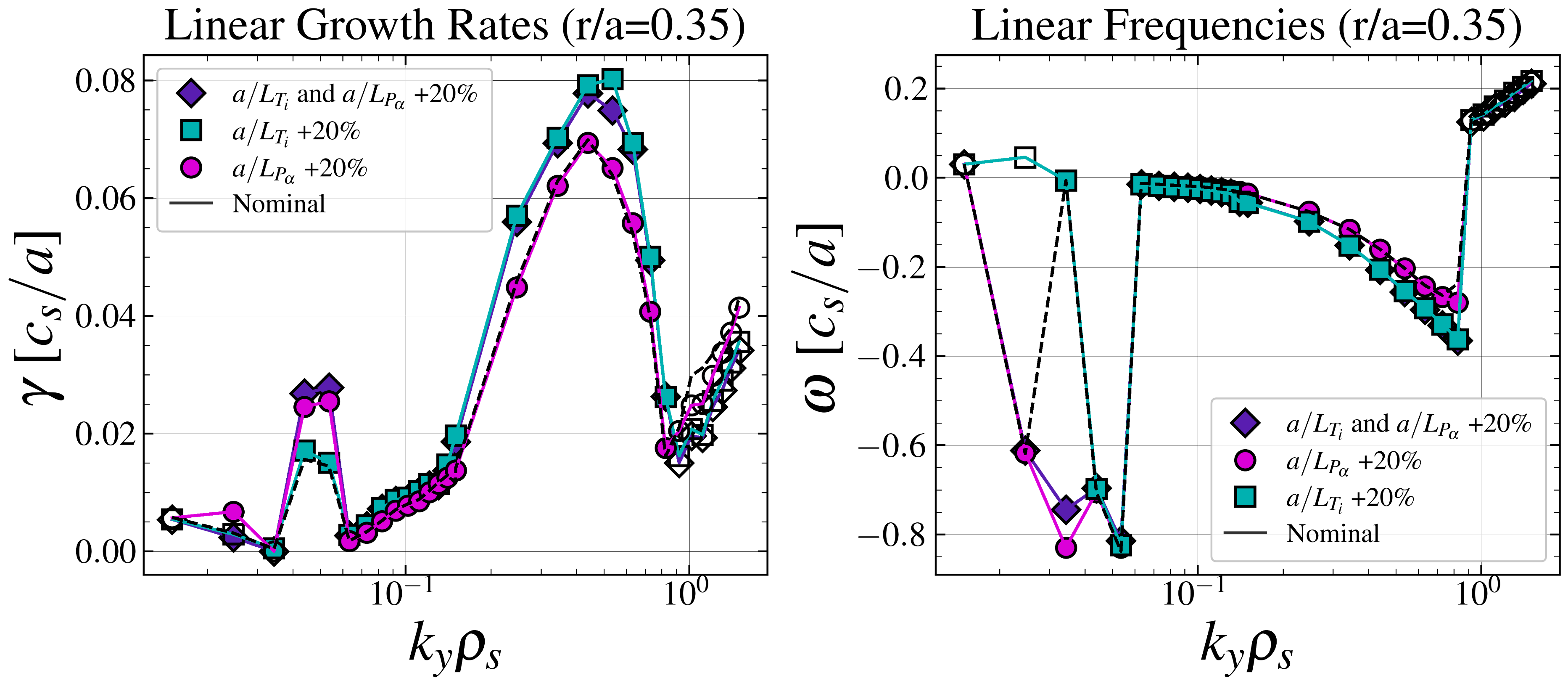}
    \caption{Sensitivity to linear growth rates to main ion temperature gradient and fast alpha pressure gradients. The ion-scale instabilities are strongly affected by a 20\% increase in $a/L_{T_i}$ but not by a 20\% increase in $a/L_{P_\alpha}$. The low-k modes are slightly affected by an increase in ion temperature gradient but strongly increased by an increase in the alpha pressure gradient. Increasing the fast ion pressure gradient destabilizes new linear modes below $k_y\rho_s\lt0.1$, but these modes are nearly stable, and are not observed in nonlinear simulations.}
    \label{fig:linearscans}
\end{figure}

\begin{figure}
    \centering
    \includegraphics[width=0.7\linewidth]{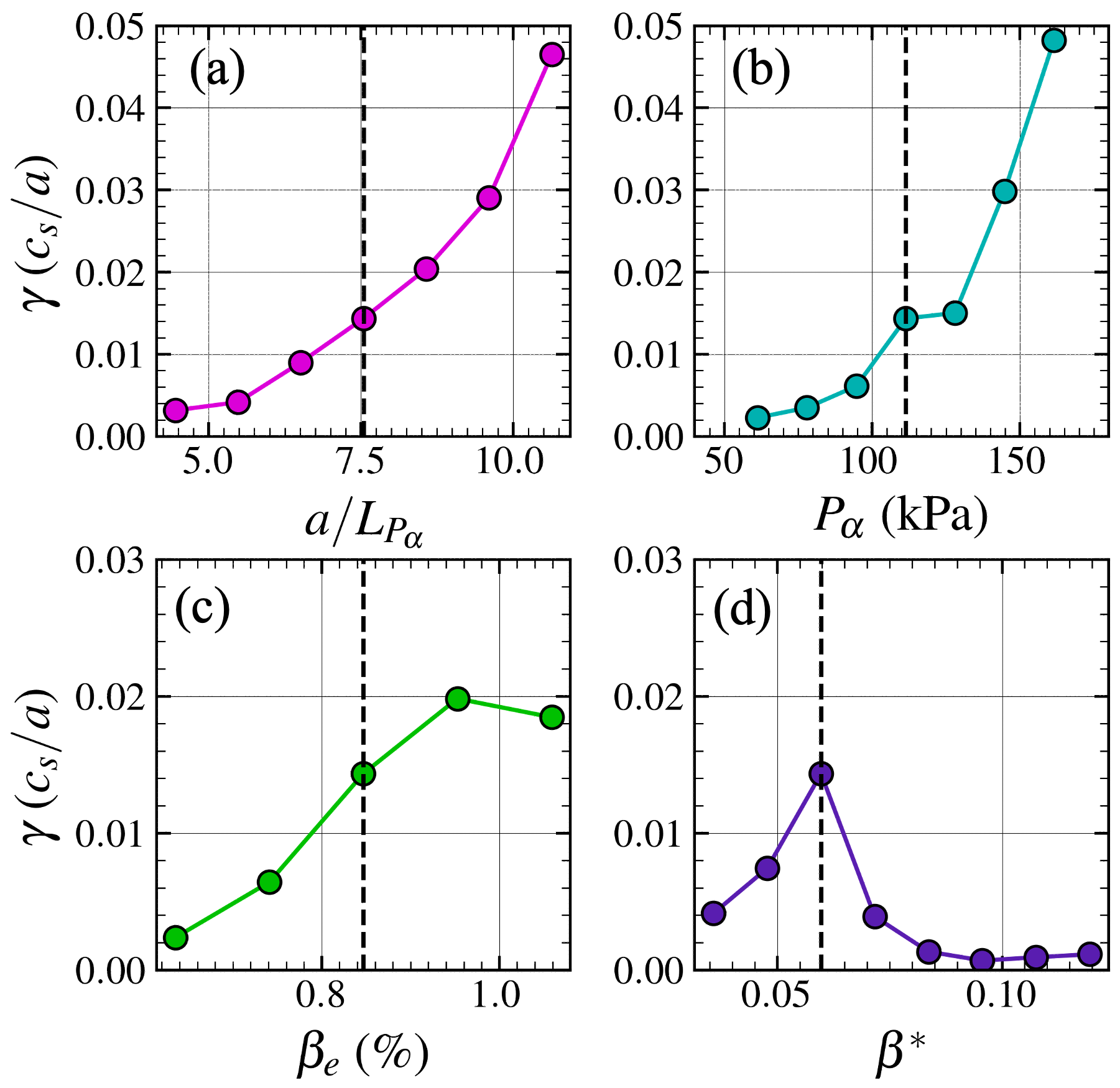}
    \caption{Sensitivity of the linearly unstable fast ion mode at $k_y\rho_s=0.0525$ to alpha pressure gradient (a), alpha pressure (b), electron beta (c), and normalized pressure gradient (d). Vertical dashed lines indicate the nominal value of each parameter.}
    \label{fig:n16sensitivity}
\end{figure}

\begin{figure}
    \centering
    \includegraphics[width=\linewidth]{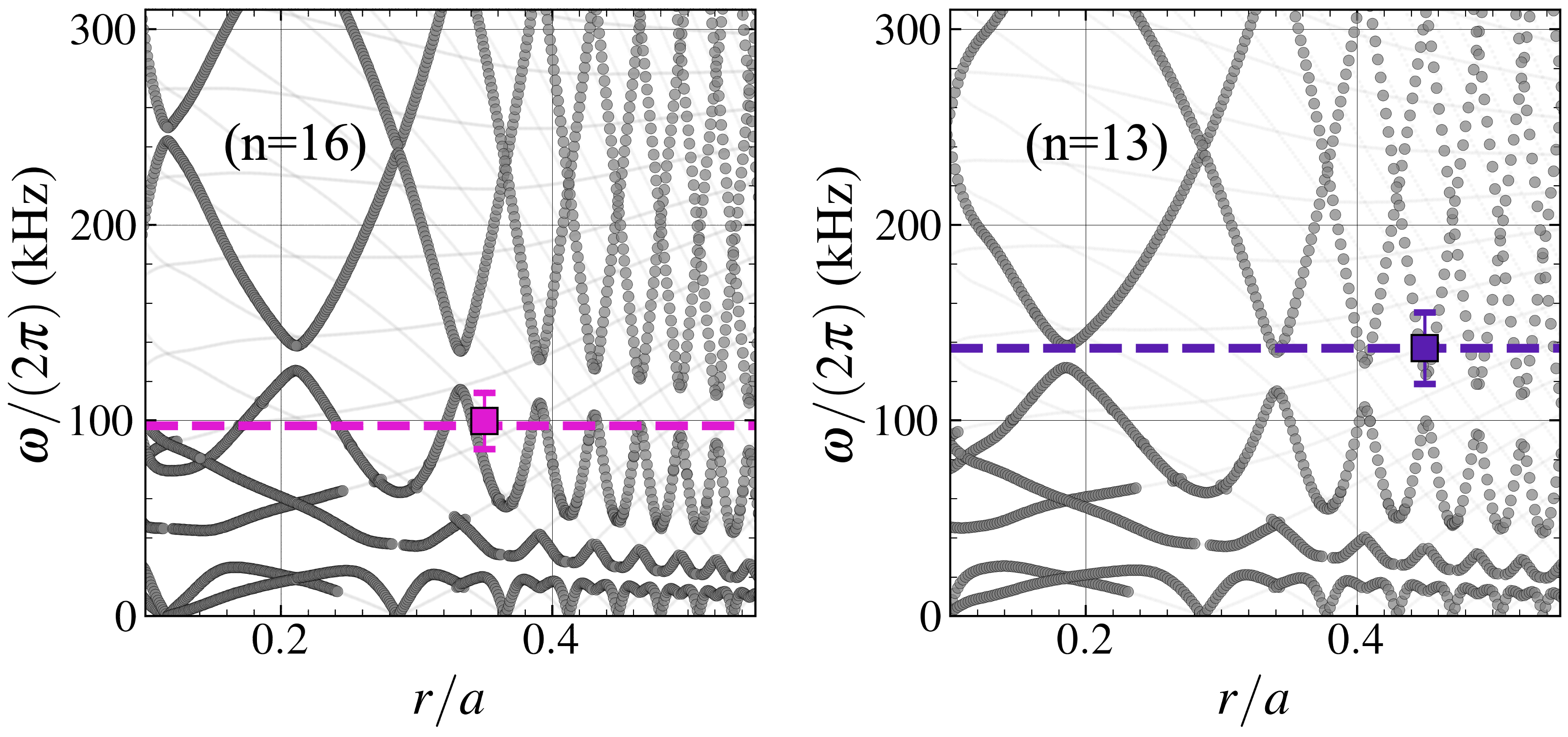}
    \caption{Shear Alfven continuum and frequency gaps for toroidal mode number $n=16$ and $n=12$ as a function of radial location $r/a$ as solved by \ALCON. Thin lines in the figures denote sound continua, and grey dots indicate Alfven continua. The linear frequency for the $n=16$ at $r/a=0.35$ mode is shown by the magenta dashed line and the nonlinear frequency (calculated in Section \ref{sec:Nonlinear CGYRO Modeling}) denoted by the marker. The $n=13$ mode at $r/a=0.45$ is shown in violet. Error bars indicate two standard deviations on the nonlinear mode frequency.}
    \label{fig:alfvencontinuum}
\end{figure}

Having examined the baseline growth rates, we next investigated the effect of altering both the thermal ion and electron gradients. as well as the fast ion pressure gradient for the simulation at $r/a=0.35$. The results of this analysis are summarized in Figure \ref{fig:linearscans}. The cases studied were with increased ion normalized temperature gradient $a/L_{T_i}$ by 20\%, fast alpha pressure gradient $a/L_{P_\alpha}$ increased by 20\% (via constant multiplication of the temperature and density gradients), and one case where both the ion temperature and fast alpha pressure gradients were increased. The increased ion temperature gradient led, unsurprisingly, to higher growth rates of ion-scale instabilities, but only a minor adjustment to the fast-ion destabilized modes at $k_y\rho_s\sim0.05$. In contrast to this, the increased fast ion pressure gradient did not significantly affect the ion-scale instabilities but resulted in a strong increase of growth rates for the low-$k$ fast ion modes. The combined effect of increasing both gradients does not change the qualitative behavior, producing only a small additive effect on the growth rates. The $k_y\rho_s=0.052$, $n=16$ fast ion mode exhibits high frequency ($\sim$100 kHz) ion diamagnetic-directed fluctuations, and is destabilized by both fast alpha pressure and pressure gradient (Figure \ref{fig:n16sensitivity}). Furthermore, its linear growth rate is increased with increasing $\beta_e$ and decreased with increasing $\beta^*$. Linear sensitivity scans show that this mode is well above marginal stability at the nominal alpha pressure gradient.
\begin{figure}
    \centering
    \includegraphics[width=0.5\linewidth]{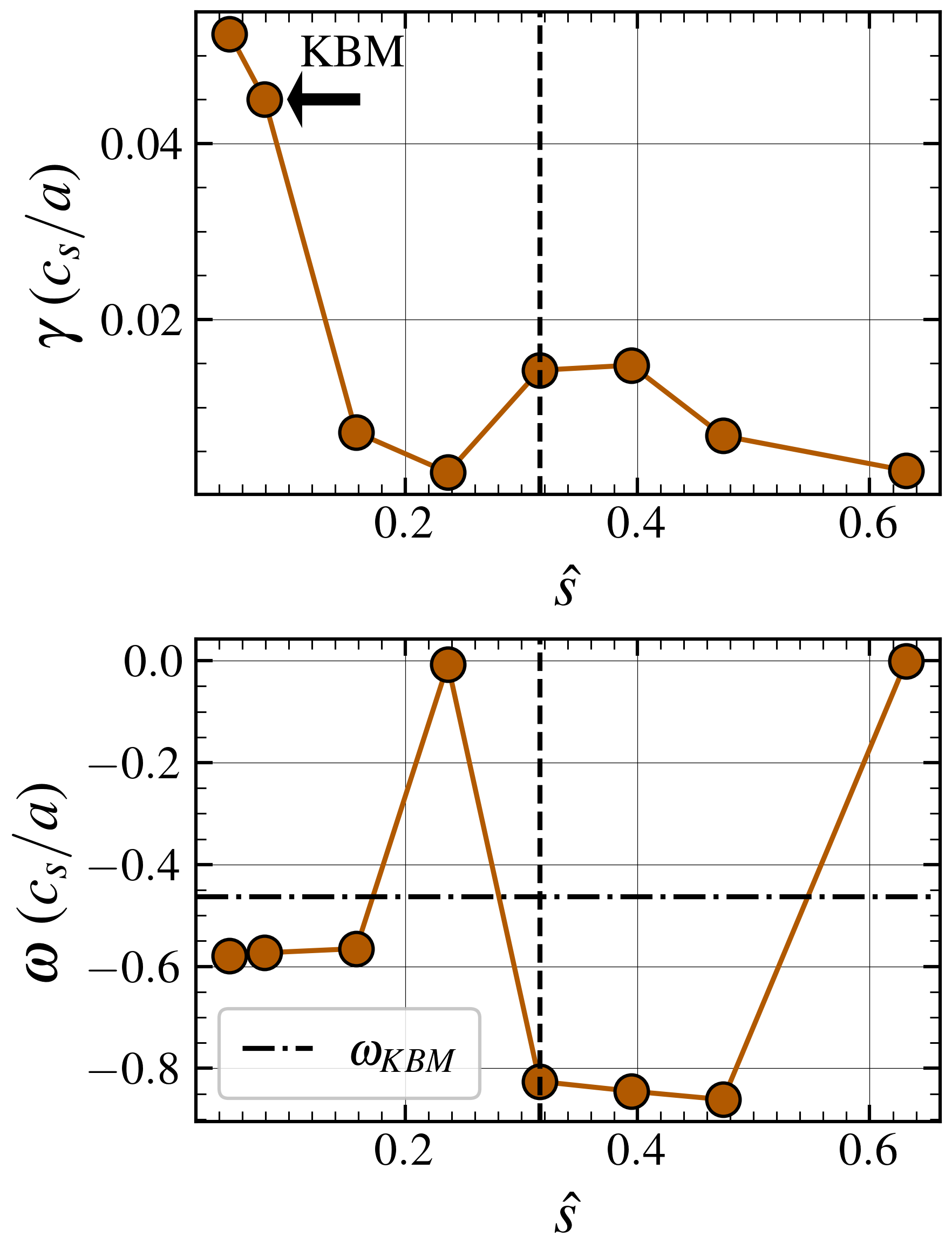}
    \caption{Linear sensitivity of the fast ion mode at $k_y\rho_s=0.0525$ to plasma shear. The mode growth rate depends non-monotonically on shear, having a linear frequency and high growth rate approximately consistent with a KBM at low shear, higher frequency consistent with a TAE at intermediate shear, and is stabilized at high shear. Vertical dashed line indicates nominal value.}
    \label{fig:shear}
\end{figure}
The presence of this mode within the range of expected unstable TAE mode numbers ($n\sim10-30$) for ARC \cite{hillesheimOverviewPhysicsBasis2026}, as well as the strong alpha particle pressure gradient and location near the $q=1$ surface suggests that it shares characteristics with the TAE. Additionally, the mode's high frequency agrees reasonably well with the analytically predicted TAE frequency gap, estimated as \cite{ajayGyrokineticInvestigationToroidal2024} $f_{TAE}\pm3\epsilon\rvert f_{TAE}\rvert/2$, where $\epsilon$ is the inverse aspect ratio and
\begin{equation}
    f_{TAE}=v_A/4\pi qR_0
\end{equation}
is the estimated TAE frequency \cite{heidbrinkBasicPhysicsAlfven2008}. The lower bound of this frequency range is $f=98$ kHz, agreeing well with the linear \CGYRO mode frequency. The mode frequency at $k_y\rho_s=0.0525$ is located near the bottom of the $n=16$ TAE frequency gap calculated with the \ALCON eigenvalue code \cite{dengLinearPropertiesReversed2012}, including additional low-frequency gaps due to finite $\beta$ compressional effects, as shown in Figure \ref{fig:alfvencontinuum}. Additionally, at $r/a=0.45$, the unstable linear mode observed at $k_y\rho_s=0.032$ ($n\sim13$) is shown to exist at the upper end of the $n=13$ TAE gap. 

\begin{figure}
    \centering
    \includegraphics[width=0.8\linewidth]{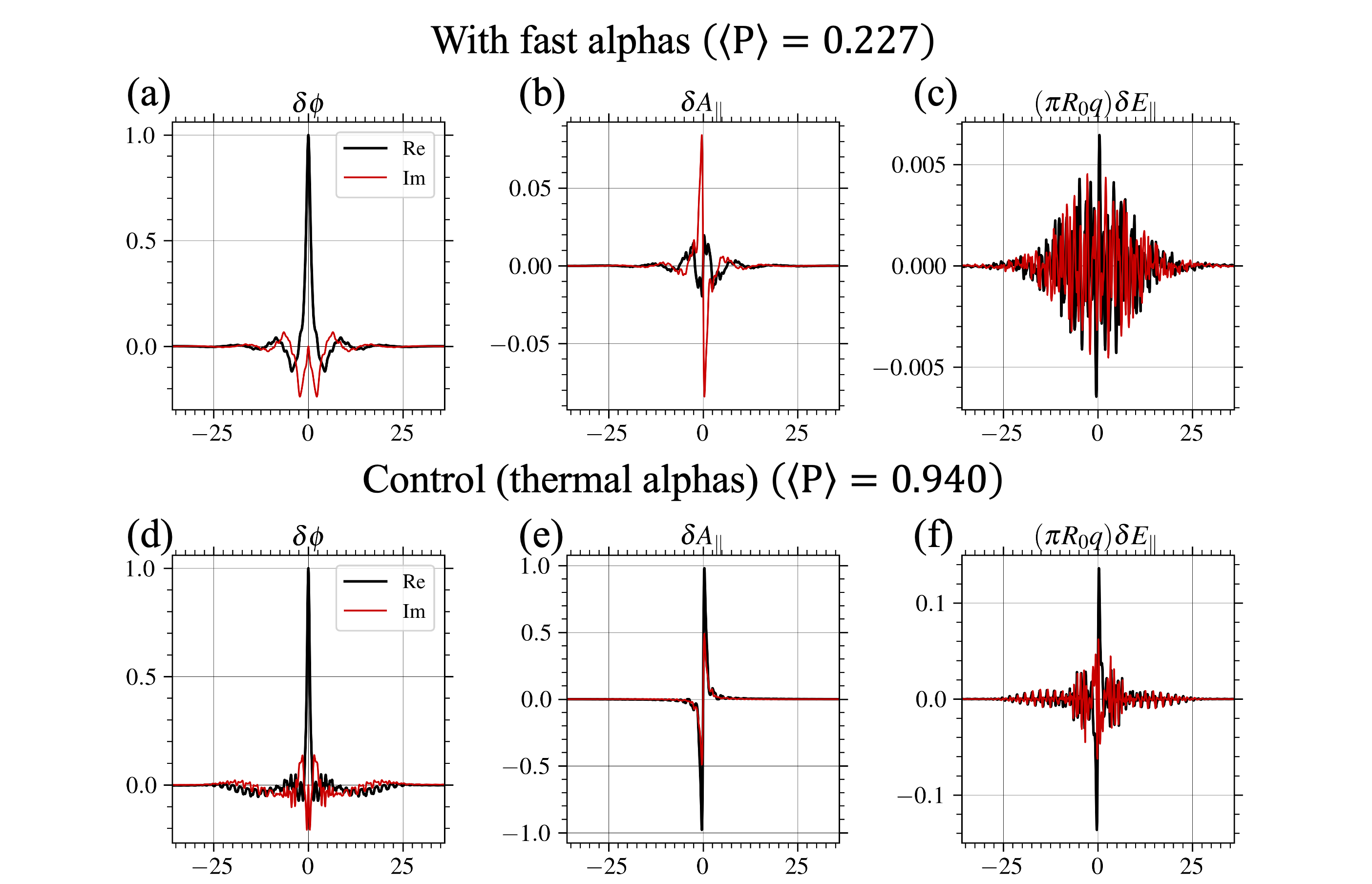}
    \caption{Linear $\delta\phi$, $\delta A_\parallel$, and $\delta E_\parallel$ fluctuations at $k_y\rho_s=0.0525$ as a function of ballooning angle with (a), (b), (c), and without fast alphas (d), (e), (f). All fluctuations are normalized to $\delta\phi_{\text{max}}$. A large $\delta E_\parallel$ is observed for the marginally stable electrostatic mode with thermal alphas, as opposed to a small but finite $\delta E_\parallel$ observed with fast alphas, quantified by the field-line averaged polarization $\langle P \rangle$.}
    \label{fig:ballooning}
\end{figure}

However, an examination of the fluctuating parallel electric field $\delta E_\parallel$ shows that the linear mode possesses a small but finite electrostatic component, unlike an ideal TAE. The mode's polarization can be derived from $\delta E_\parallel=-ik_\parallel \delta\hat{\phi}+i\omega \delta \hat{A}_\parallel/c$ by introducing
\begin{equation}
    P = \frac{\rvert\delta \hat{\phi}\rvert}{\rvert\delta \hat{\phi}\rvert
    +\rvert\delta \hat{\psi}\rvert}
\end{equation}
where $\delta\psi=\omega \delta\hat{A}_\parallel /ck_\parallel$. Defined in this way, a purely electrostatic mode approaches $P\sim1$, while for a shear Alfven wave, $E_\parallel=0$ by necessity, giving $P\sim0.5$. The ballooning-space eigenvalue structure and polarization for the mode at $k_y\rho_s=0.0525$ with and without fast alphas is shown in Figure \ref{fig:ballooning}, demonstrating the inherently electromagnetic character of the fast ion mode. The field line-averaged polarization for the fast ion mode is $\langle P \rangle=0.227$, showing imperfect cancellation of electromagnetic and electrostatic field components, compared to the dominantly electrostatic mode observed without fast alphas, with $\langle P \rangle=0.940$. The mode was not observed when $\delta B_\parallel$ fluctuations were neglected in linear simulations, confirming that compressional magnetic effects are important to include in both linear and nonlinear simulations to capture the full physics of fast alpha-driven instabilities. 
A finite parallel electric field arising from the kinetic fast ion response was predicted for Alfven waves as early as \cite{rosenbluthExcitationAlfvenWaves1975}. It has since been observed experimentally that the TAE accumulates a significant $\delta\phi_\parallel$ when strong suppression of ITG turbulence by TAE-driven zonal flows is occurring \cite{duMicroturbulenceSuppressionAlfven2025}. This nonlinear upshift in polarization is understood theoretically to be the driving mechanism of TAE-enhanced zonal flows, as $\delta\phi_{ZF}\sim ik_x\left\rvert \delta\phi_{\text{TAE}} \right\rvert^2$ \cite{qiuNonlinearExcitationFiniteradialscale2017,chenEffectsZonalFields2024}. A more careful analysis of the polarization of fast alpha-driven modes in nonlinear simulations and comparison to experiment is thus a promising subject for further work.

Finally, a linear scan in shear (Figure \ref{fig:shear}) reveals that the mode becomes briefly stable, then becomes strongly unstable as  shear is decreased, with the linear frequency approaching the analytically predicted Kinetic Ballooning Mode (KBM) frequency \cite{maTheoreticalStudiesLowfrequency2022}
\begin{equation}
    \omega_{\text{KBM}}=\sum\omega^*_{pi}f_i
\end{equation}
where $\omega^*_{pi}=k_y\rho_s(T_i/T_e)(a/L_{Pi})(c_s/a)$ is the diamagnetic frequency and $f_i=Z_in_i/n_e$ the fraction of ion species $i$. For the profiles investigated here the effective KBM frequency is $\omega_{\text{KBM}}\sim54$ kHz. This analysis raises the possibility that KBM-like fast ion modes similar to those observed in \cite{kimInvestigationFastIon2025a} could be observed at lower shear, where existing work has already demonstrated that KBM-driven electromagnetic turbulence limits core transport in 
the full-power ARC V3A scenario \cite{howardPerformanceTransportARC2026}.

\section{Nonlinear \CGYRO Modeling}
\label{sec:Nonlinear CGYRO Modeling}

In this section we cover the main results of this paper from nonlinear \CGYRO simulations. This section proceeds as follows: first, the nonlinear simulation setup is described. Next, we examine the $r/a=0.35$ location, observing strong turbulence suppression effects associated with a multi-scale coupling between fast alpha-driven instabilities and ion-scale turbulence which results in saturated values of the energy and particle fluxes several times less than the control simulation. Next, the dependence of heat and particle fluxes on the normalized ion temperature, electron temperature, and electron density gradients is studied, revealing that the reduction in heat flux is best explained by an upshift in the critical gradient due to the highly driven zonal flows. An investigation at other radial locations reveals that the nonlinear stabilization effects are limited to the innermost radii for which the alpha density is high. The rest of the section is devoted to a sensitivity study of the fast alpha density and density gradient and $\beta_e$, both of which reveal the fundamentally electromagnetic character of turbulence stabilization by fast ion mode -- zonal flow -- ITG interaction and the beneficial scaling of this effect with alpha pressure.

\subsection{Simulation setup}

Nonlinear \CGYRO simulations were performed across a range of radial locations and normalized temperature and density gradients. All simulations were, unless otherwise stated, performed over a range of binormal wavenumbers $k_y\rho_s=[0.0263, 1.234]$, chosen to accurately resolve fluctuations up to toroidal mode $n\sim8$ at the innermost radii, down to the scale of the tritium sound speed gyroradius $\rho_T=\sqrt{m_T/m_D}\rho_s$, where $\rho_s$ is the deuterium sound speed gyroradius 
\begin{equation}
    \rho_s \dot{=} \frac{c_s}{eB_{unit}/(m_Dc)}
\end{equation}
used for normalization in \CGYRO \cite{candyHighaccuracyEulerianGyrokinetic2016}. This corresponds to 48 individual toroidal modes per simulation, unless otherwise specified. The number of radial grid points for all simulations was $N_r=512$, resulting in a minimum radial scale resolution of $k_x\rho_s=6.64$, sufficiently resolving ion-scale fluctuations in the radial direction. At $r/a=0.35$, with the local values of the $q$-profile and shear this corresponds to a box size of $[L_x,L_y]=[241\rho_s,239\rho_s]$ according to $L=2\pi/k_{y,0}\rho_s$, resolving toroidal mode numbers from $n=8$ to $n\sim375$. Both control and fast ion simulations were performed using the same box sizes and resolution for consistency. Nonlinear simulations used 24 points of resolution in both pitch angle $\xi$ and poloidal angle $\theta$, and 8 points in energy. Box size convergence testing was performed for nonlinear simulations, and simulations were considered resolved when time-averaged turbulent fluxes between resolutions did not differ by more than two standard deviations. At $r/a=0.35$, nonlinear simulations were tested down to a minimum $k_{y}\rho_s=0.016$ $(n=5)$ using 64 toroidal modes, without significant change in the saturated fluxes. Convergence testing in poloidal resolution was also performed, where increasing to $N_\theta=32$ did not produce a significant change in turbulent fluxes. All simulations presented here used the Sugama collision operator \cite{sugamaLinearizedModelCollision2009}, a high-fidelity model including all species collisions and appropriately modeling collisional damping of the zonal flow rate \cite{belliImplicationsAdvancedCollision2017}. Simplified collision operators such as the fast diagonal Lorenz model included in \CGYRO are not appropriate for capturing the saturated turbulence fluxes in the presence of fast ions for the cases studied, as such models neglect off-diagonal collisions which significantly affects the collisional damping of zonal flows \cite{belliImplicationsAdvancedCollision2017}, and lead to unphysically low heat fluxes. Off-diagonal terms in the collision operator were represented with 32-bit precision, effectively halving the memory required while producing no noticeable change in the resulting heat fluxes. In addition to electrostatic potential $\phi$, both perpendicular and parallel electromagnetic fluctuations $A_\parallel$ and $B_\parallel$ were simulated. Simulations varied widely in the amount of time needed for turbulent fluxes to reach a steady-state. All simulations were investigated individually to ensure proper saturation had been reached. Heat and particle fluxes reported in this work were averaged over the last half of the simulation time, or by at least $500$ $a/c_s$ for runs shorter than $1000$ $a/c_s$. The standard error for the flux channel $Q$ is defined as $\Delta Q=\sigma_Q/\sqrt{N_{\text{eff}}}$ using an effective sample size determined by $N_{\text{eff}}=N/3\tau_c$. In this expression, $N$ is the total number of simulation time points, and $\tau_c$ is the $e$-folding autocorrelation time, defined by $\tau_c=\arg \min_\tau \rvert R(\tau)-e^{-1} \rvert$, with the autocorrelation function $R$ defined for a time series with mean $\mu$ and variance $\sigma^2$ in the standard way:
\begin{equation}
\hat{R}_{XX}(\tau)=\frac{1}{(N-\tau) \sigma^2} \sum_{t=1}^{N-\tau} [X(t)-\mu][X(t+\tau)-\mu]     
\end{equation}

Using this formula for the effective sample size thus ensures that uncertainties reported reflect the number of independent samples within the averaging window. Nonlinear simulations including fast alpha particles were performed on AMD MI300A APU nodes at the San Diego Supercomputing Center COSMOS cluster, while control simulations were performed on NVIDIA A-100 GPU nodes on NERSC's Perlmutter system. Additional convergence testing was performed on NVIDIA H200 GPU nodes on MIT's Engaging Cluster. This work benefited in particular from recent GPU memory offload techniques implemented in \CGYRO, allowing the very large memory requirement for the collision operator to be reduced by another factor of 2.

\subsection{Nonlinear turbulence suppression at $r/a=0.35$}

\begin{figure}
    \centering
    \includegraphics[width=\linewidth]{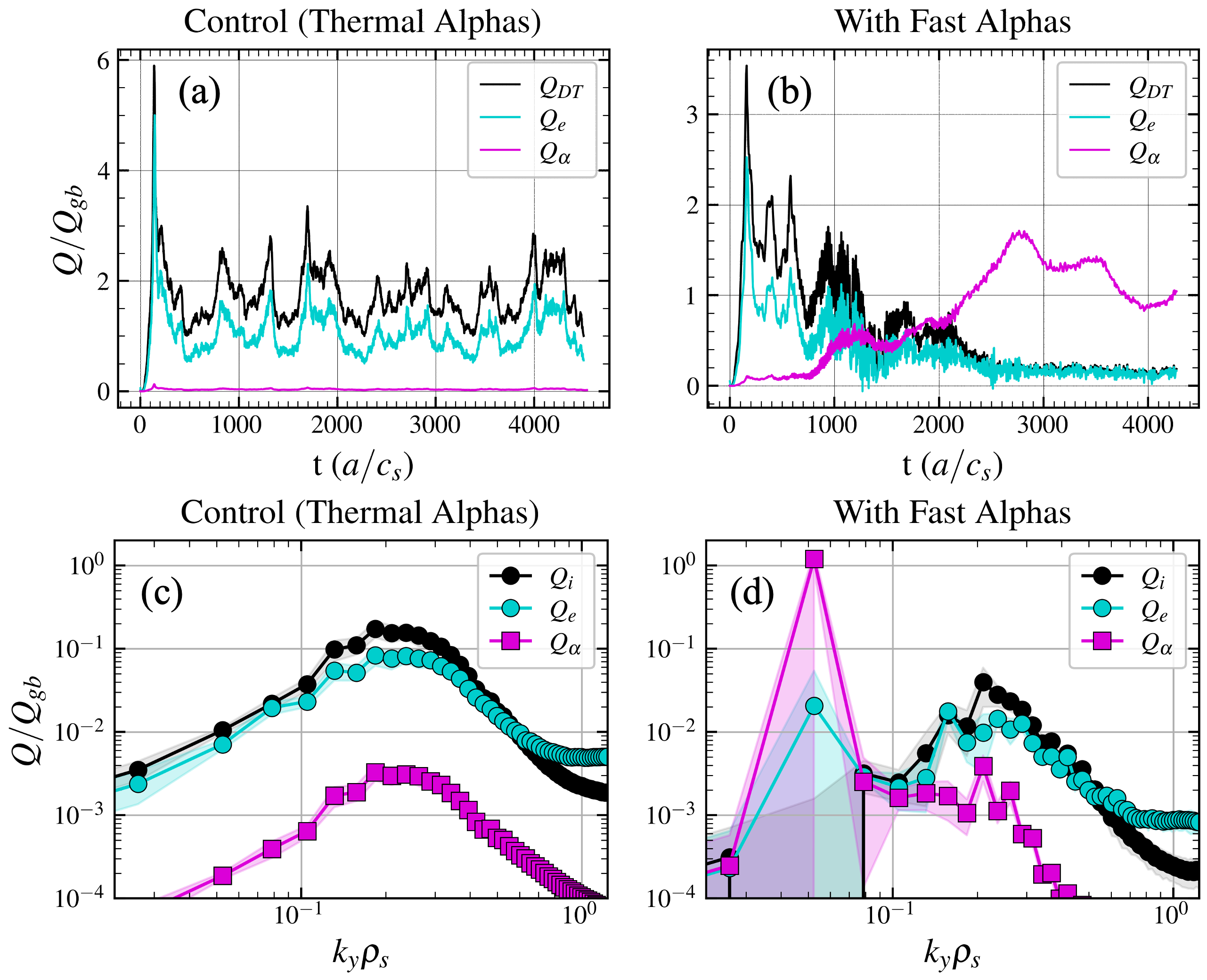}
    \caption{Nonlinear heat flux in gyro-Bohm units and heat flux spectra for the \CGYRO simulation at $r/a=0.35$ with (b, d) and without (a, c) fast alphas. Significant reduction in both main ion and electron heat fluxes in the presence of fast alphas (b) is observed along with high alpha particle heat fluxes. The ion heat flux spectrum demonstrates that this increase in alpha heat flux is localized to the unstable fast ion mode at $k_y\rho_s=0.0525$. Shaded areas in the heat flux spectrum (d) reflect the 95\% confidence interval for the mean flux at that wavenumber.}
    \label{fig:standalone_timetraces_and_fluctuations}
\end{figure}

\begin{figure}
    \centering
    \includegraphics[width=0.9\linewidth]{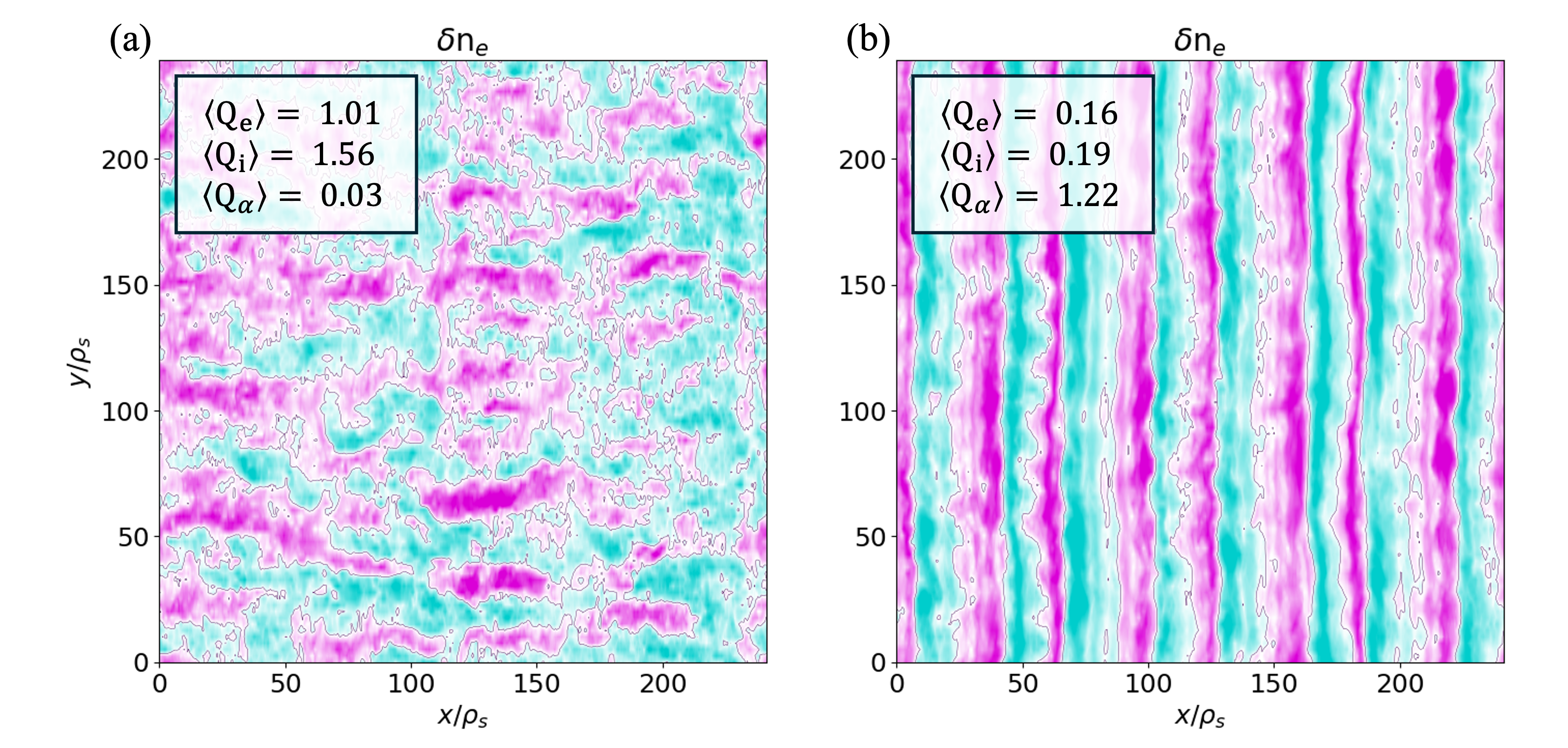}
    \caption{Real-space density fluctuations for control (a) and fast alpha (b) simulations at $r/a=0.35$ at $t=4000$ $a/c_s$. Compared with the control simulation, the enhanced zonal flow shearing rate associated with the fast ion mode creates a sheared pattern of isolated turbulent eddies, along with dramatically reduced heat flux.}
    \label{fig:035_density_fluct}
\end{figure}

Nonlinear simulations at $r/a=0.35$ were performed at the nominal plasma gradients predicted with medium-fidelity modeling, as shown in Figure \ref{fig:medfid}. The time traces of the control and fast ion simulations are presented in Figure \ref{fig:standalone_timetraces_and_fluctuations}. In addition, a nonlinear simulation was conducted with the normalized alpha pressure gradient artificially set to zero. The presence of fast alpha particles, though having a minor impact on linear growth rates, was observed to significantly impact both the saturated heat fluxes and fluctuation amplitudes. As has been observed in previous work \cite{disienaElectromagneticTurbulenceSuppression2019}, a characteristic two-phase behavior is seen in the time trace of main ion heat fluxes. In the first phase, the heat flux oscillates rapidly. This period of rapid oscillation in the heat flux that correlates with the increase in amplitude of the potential fluctuation at $k_y\rho_s=0.0525$ and the alpha heat flux, which is also plotted in Figure \ref{fig:standalone_timetraces_and_fluctuations}. After the fast ion instability saturates, the second phase of the nonlinear simulation is characterized by significantly suppressed turbulent heat flux relative to the control simulation, with a mean ion heat flux of $\langle Q_{DT}^c/Q_{GB}\rangle =1.601$ in the control simulation compared with $\langle Q_{DT}^f/Q_{GB}\rangle =0.171$ with fast alpha particles. The heat flux spectra in Figure \ref{fig:standalone_timetraces_and_fluctuations} additionally demonstrate that the fast particle heat flux is localized in wavenumber to $k_y\rho_s=0.0525$, which corresponds to the high-frequency fast ion mode observed in linear simulations. The real-space electron density fluctuations are shown in Figure \ref{fig:035_density_fluct}, in which the strongly sheared zonal structures in the binormal direction are visible, along with significantly reduced turbulent fluctuations in the radial direction. In contrast, although some zonal structure is visible in the control simulation, the density fluctuations are of increased amplitude with eddies extended in the radial direction. Saturated ion heat flux was moderately lower in the case with zero normalized alpha pressure gradient, with an average ion heat flux value of $\langle Q_{DT}/Q_{GB}\rangle =1.152$. The difference compared to the control simulation is likely due to the ``direct" ITG stabilization effect due to the fast alpha contribution to $\beta_i$. The remaining discrepancy in heat flux is a fully nonlinear effect associated with the fast alpha-driven modes. As expected, while direct stabilization effects are present, they are not the dominant mechanism.

To test the hypothesis that the turbulence suppression is caused by the fast ion instability-generated zonal flows, we introduce the total zonal drift energy \cite{belliFlowshearDestabilizationMultiscale2024} as a measure of the zonal flow shearing rate, defined by

\begin{equation}
\label{eq:k0}
    K^{(0)}_{tot}\doteq \sum_{k_x\rho_s=-1}^{1}k_\perp^2\rho_s^2\left\langle \abs{\hat{\phi}(k_y,k_x)}^2 \right\rangle\biggr\rvert_{k_y=0}
\end{equation} 

\begin{figure}
    \centering
    \includegraphics[width=\linewidth]{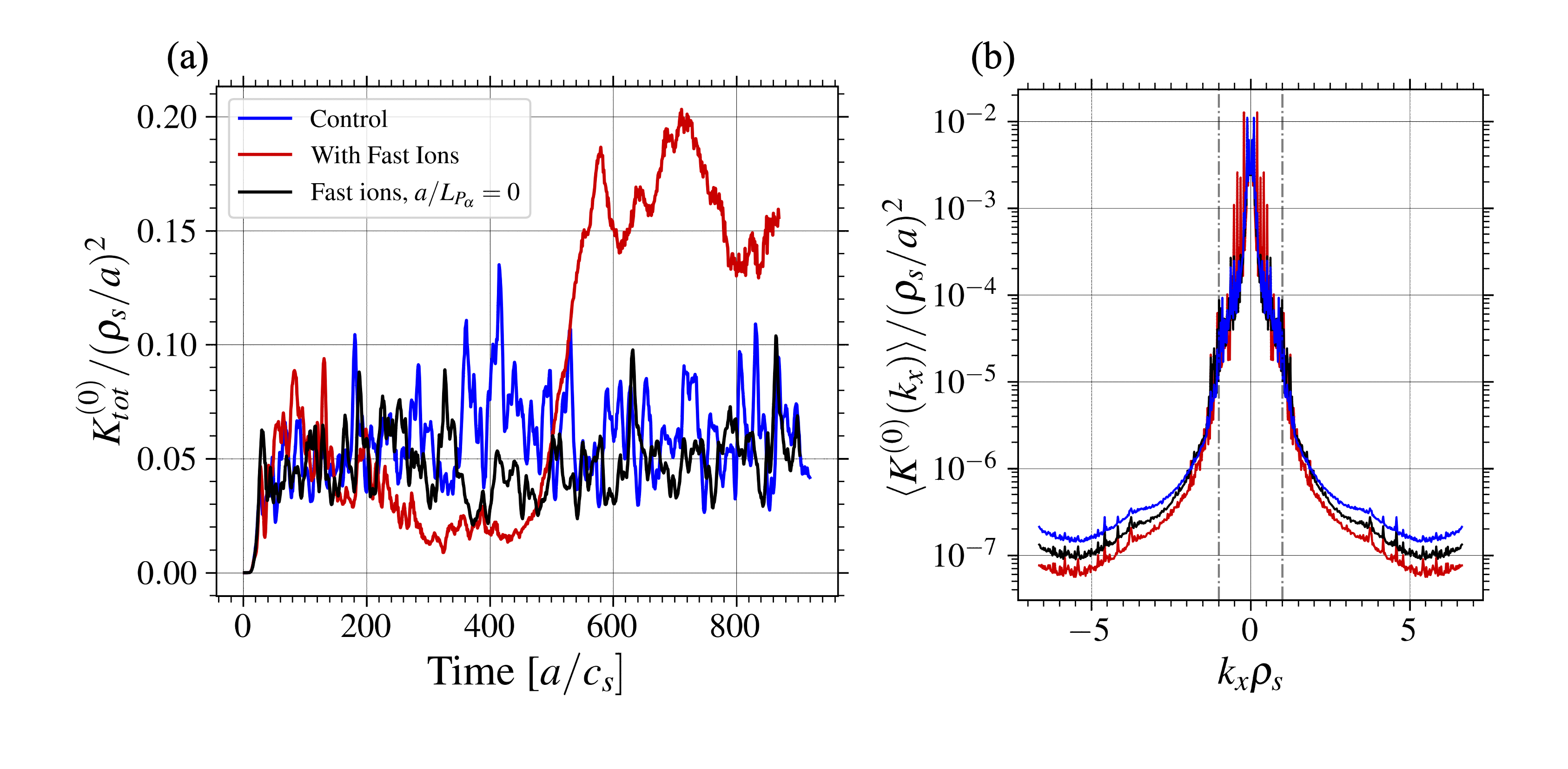}
    \caption{(a) total zonal flow drift energy normalized to $\rho_s^2/a^2$ for nonlinear control, fast alpha, and fast alpha with artificially zeroed alpha pressure gradient simulations. While the case with fast ions but no pressure gradient has slightly lower heat flux than the control simulation, the zonal shearing rate does not change, indicating that zonal flow stabilization effects are not present. (b) $k_x$-spectrum of zonal flow drift energy, averaged over time and $k_y$. Dashed lines indicate the limits of summation in eq. \eqref{eq:k0}, demonstrating that small-scale fluctuations do not contribute significantly to the zonal flow shearing rate. In particular, the high value of $K^{(0)}$ for the simulations with large scale unstable fast alpha-driven modes is apparent at very low radial wavenumber.}
    \label{fig:zonal}
\end{figure}

Where $k_\perp^2$ is approximated as $k_\perp^2\approx k_y^2+k_x^2$. The $k_x\rho_s=0$ contribution is neglected, as this component of the zonal field does not contribute to energy transport. The total zonal drift energy as a function of time for the three nonlinear simulations is plotted in Figure \ref{fig:zonal}(a). Additionally, the radial wavenumber spectrum of $K_{tot}^{(0)}$ is shown in Figure \ref{fig:zonal}(b), demonstrating that the primary contribution to the zonal drift energy is indeed from the ion-scale modes.
$K_{tot}^{(0)}$ is observed to fluctuate in the initial phase, but rises to a higher saturated level in the case with fast alphas, corresponding to the saturated state of the heat flux beyond $t\approx2500a/c_s$ in Figure \ref{fig:standalone_timetraces_and_fluctuations}. The increase in the zonal flow shearing rate closely follows the growth of the potential fluctuation $\delta\phi(k_y\rho_s=0.0525)$, supporting the view that this enhanced zonal flow drive is connected to the instability growth and its nonlinear saturation. 

Next, we report the results of a sensitivity study of the normalized ion and electron temperature and electron density gradients. The observed reduction in heat flux at $r/a=0.35$ would be unlikely to translate to increased fusion performance unless a higher normalized gradient can be accessed at a constant heat flux, relative to the control case. The shape of the flux-gradient relationship for a given radial location therefore determines whether a given profile can be steady-state. The power balance, or ``target" flux is used here to give context to these nonlinear results with a view towards future studies looking at overall device performance. As the local gradients and power balance is calculated using medium-fidelity \MAESTRO-\TGLF simulations, it is not necessarily expected that \CGYRO results will be flux-matched for the profiles used here. Whether the saturated turbulent heat or particle fluxes from \CGYRO are higher or lower than the targets is an indication of how the profile \textit{would} evolve if brought to steady-state. 

\begin{figure}
    \centering
    \includegraphics[width=\linewidth]{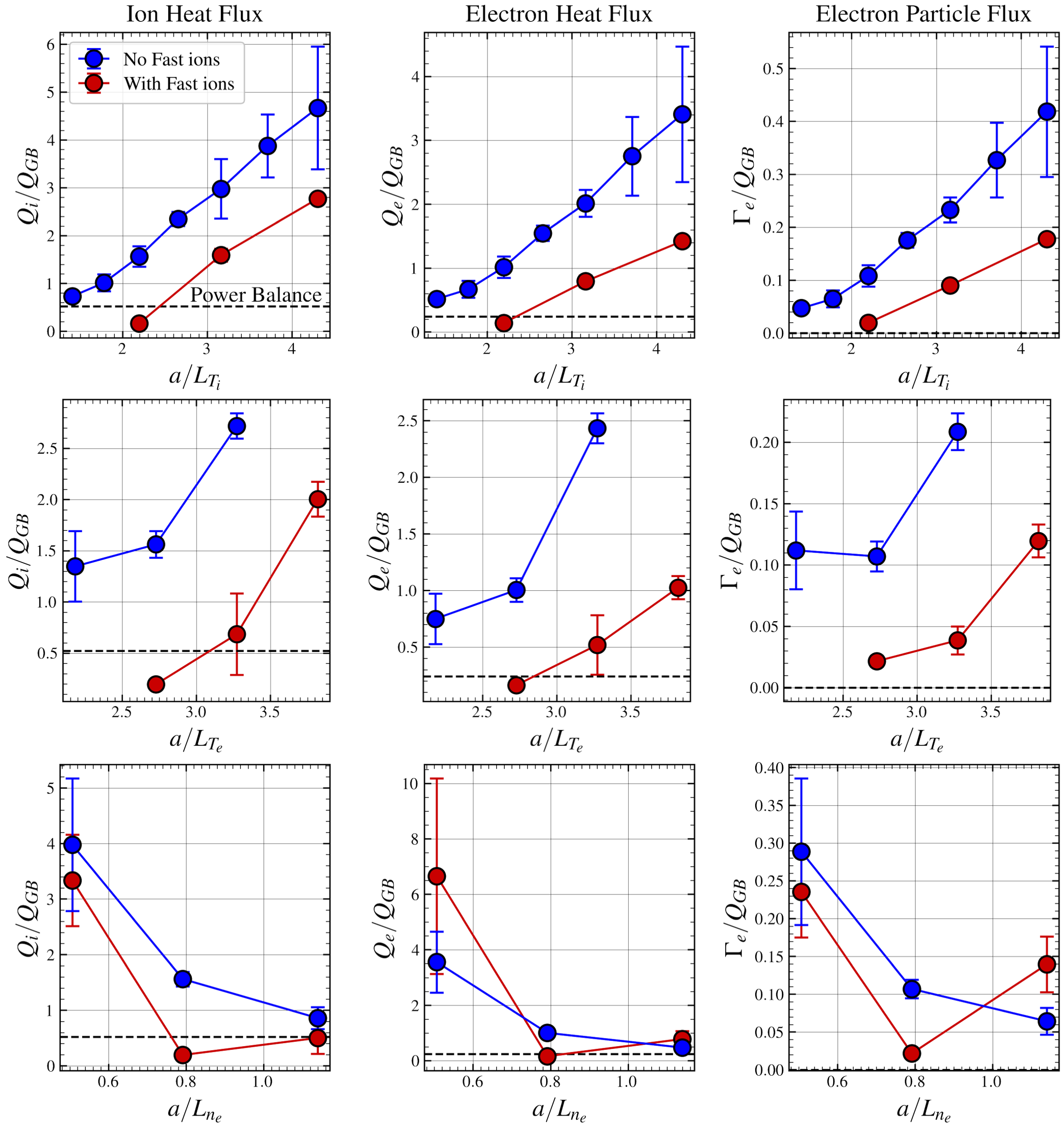}
    \caption{Scans of normalized main ion temperature gradient $a/L_{T_i}$, electron temperature gradient $a/L_{T_e}$, and electron density gradient $a/L_{n_e}$. Dashed line labeled ``Power Balance" denotes the volume-averaged sources and sinks of heat and particles inside of $r/a=0.35$ predicted with \MAESTRO.}
    \label{fig:gradientscan}
\end{figure}

A nonlinear analysis of the flux-gradient relationship at $r/a=0.35$ is presented in Figure \ref{fig:gradientscan}. This analysis reveals two important features of the observed turbulence suppression. Firstly, the ITG turbulent heat flux is lower at all points in the ion temperature gradient scan, relative to the control simulations. Secondly, the stiffness, defined as $\partial (Q_{s,GB})/\partial (a/L_{T_s})$ is only weakly affected by the inclusion of fast alpha particles. The simulations including fast alphas show a nonlinear upshift in the critical gradient, consistent with the physical picture of a fast ion instability-driven zonal flow being the mechanism for turbulence suppression \cite{dimitsComparisonsPhysicsBasis2000,garbetGyrokineticSimulationsTurbulent2010}. This is a promising result if this relationship holds for steady-state burning plasmas, as strongly driven ITG turbulence and high gyro-Bohm unit fluxes is expected to pin temperature profiles at the ITG critical gradients \cite{hollandDevelopmentCompactTokamak2023,howardHighFidelityPredictions2025}, making an upshift in the critical gradient from fast alpha particle effects likely more relevant for increasing fusion performance compared to a reduction in stiffness. The increase in ion temperature gradient at a constant heat flux for this radial location is nearly 25\%. One question raised by these results is how an enhanced Dimits shift-type effect is possible without the zonal modes being limited by a tertiary instability \cite{rogersNoncurvaturedrivenModesTransport2005}. One possible explanation is that compared to ITG-scale zonal fluctuations, TAE and other low-$k_x$ modes produce longer-wavelength zonal flow oscillations which have comparatively low curvature $\mathrm{d}^2v_{ZF}/\mathrm{d}r^2$, which has been identified as the driving term for tertiary instabilities which limit zonal flows \cite{zhuTheoryTertiaryInstability2020}.

\begin{figure}
    \centering
    \includegraphics[width=\linewidth]{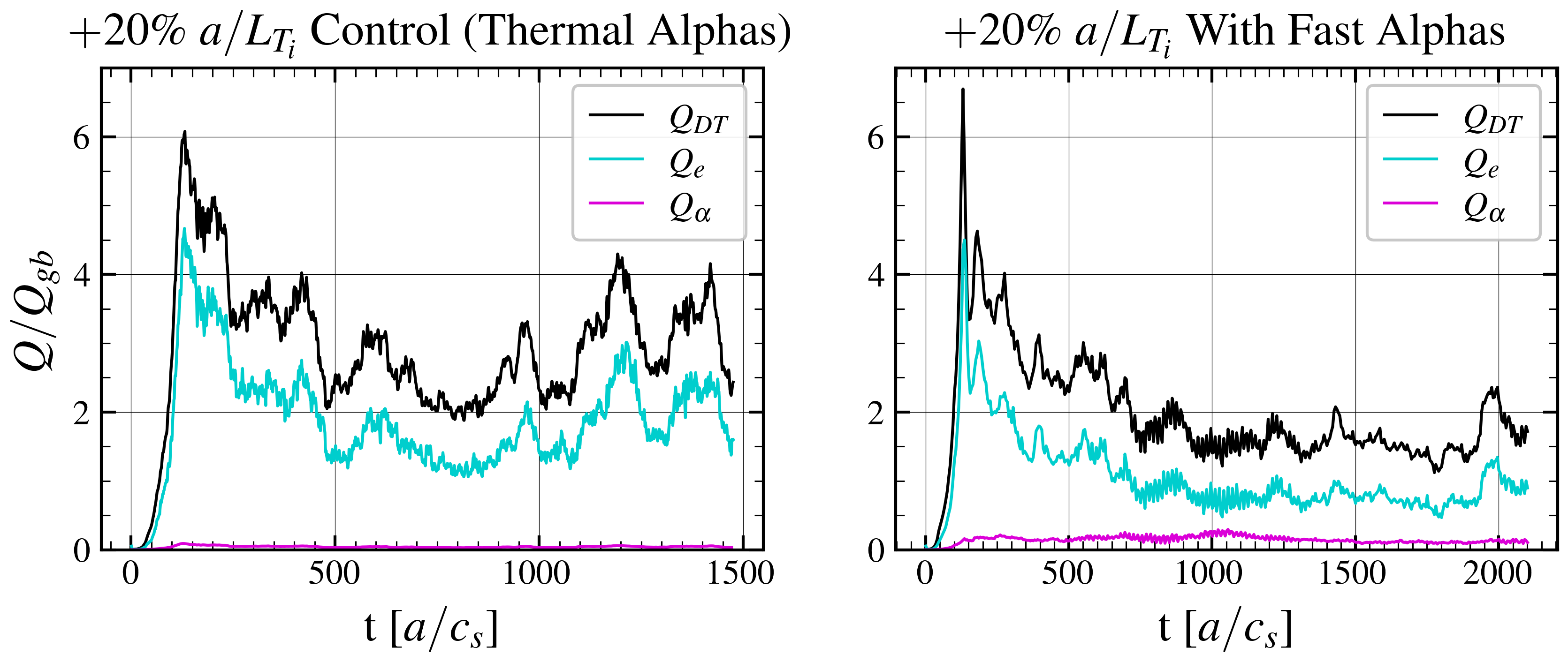}
    \caption{Nonlinear heat fluxes at $r/a=0.35$ with 20\% increased $a/L_{T_i}$. Despite significantly lower alpha heat fluxes, the high-frequency oscillation in the heat flux near $t=1000$ $a/c_s$ is still present, and the saturated heat flux remains lower than in the control simulation. A moderately high alpha heat flux from the unstable mode is also present.}
    \label{fig:increasedaltiflux}
\end{figure}

The extended zonal structures observed in the nonlinear simulation with fast alpha particles are likely unphysical, due to the gradients predicted by \MAESTRO being too low at this location to strongly drive ITG. This can be seen in Figure \ref{fig:gradientscan}, as the nominal gradients produce electron and ion heat fluxes below the power balance level. These zonal structures disappear at higher electron and ion temperature gradients. For instance, at +20\% $a/L_{T_i}$ the nonlinear density fluctuations do not exhibit box-size structures, even as the nonlinear heat fluxes are reduced by a factor of two relative to the control simulation (Figure \ref{fig:increasedaltiflux}). In light of these observations, the subsequent analyses (alpha particle density and density gradient, $\beta_e$) were performed with +20\% increased $a/L_{T_i}$. In this simulation, the fast alpha-driven mode is seen to to contribute significantly to stabilization despite strongly driven ITG turbulence.

\subsection{Scans in alpha particle density and pressure gradient}
\label{subsec:alphascans}

\begin{figure}
    \centering
    \includegraphics[width=0.8\linewidth]{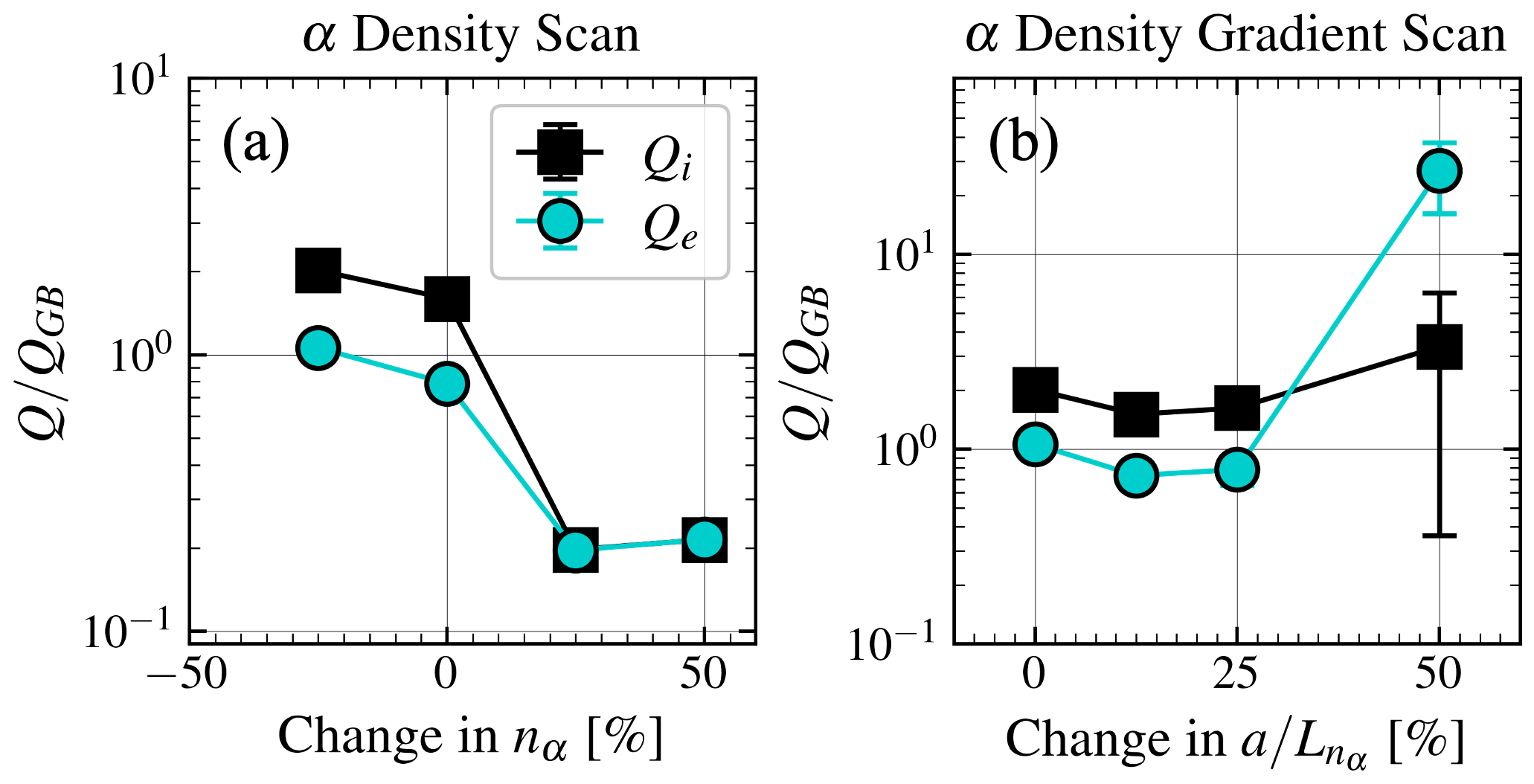}
    \caption{Scans in alpha particle density (a) and density gradient (b) at $r/a=0.35$ with $a/L_{T_i}$ +20\% relative to \MAESTRO profiles. (a) increasing alpha particle density leads to a drop in both main ion and electron heat flux, while decreasing alpha density leads to a modest increase in heat flux. Both of these behaviors are consistent with the picture of turbulence suppression by fast ion destabilized modes. (b) Modest heat reduction change is observed with increases up to 25\% in alpha density gradient, however increasing the density gradient by 50\% strongly destabilizes the $n=16$ fast alpha-driven mode and leads to a strong increase in heat flux along with a large increase in alpha particle flux.}
    \label{fig:alphadensity}
\end{figure}

In this section, we investigate the effects of changing the alpha particle density gradient, along with the alpha particle density. Both of these parameters affect the stability of the fast alpha-destabilized mode, which has a linear growth rate that scales as $\gamma\propto\beta_\alpha f_\alpha R_0/L_{P_\alpha}$ \cite{zoncaResonantDampingToroidicityinduced1992, zoncaTheoryContinuumDamping1993}, where $\beta_\alpha$ is the alpha particle beta, $f_\alpha=n_\alpha/n_e$, and $L_{P_\alpha}$ is the alpha pressure gradient scale length. Here, only the alpha particle density gradient is changed, due to the observation that the normalized temperature gradient for the alpha particles does not increase with simple collisional slowing-down transport, even at higher fusion power. This behavior was confirmed by additional high-fidelity \NUBEAM simulations. Results from nonlinear simulations in Figure \ref{fig:alphadensity} demonstrate that increased alpha density has a large impact on the observed heat flux, reducing the heat flux to a similar value observed with the nominal $a/L_{T_i}$.

\begin{figure*}
    \centering
    \includegraphics[width=\linewidth]{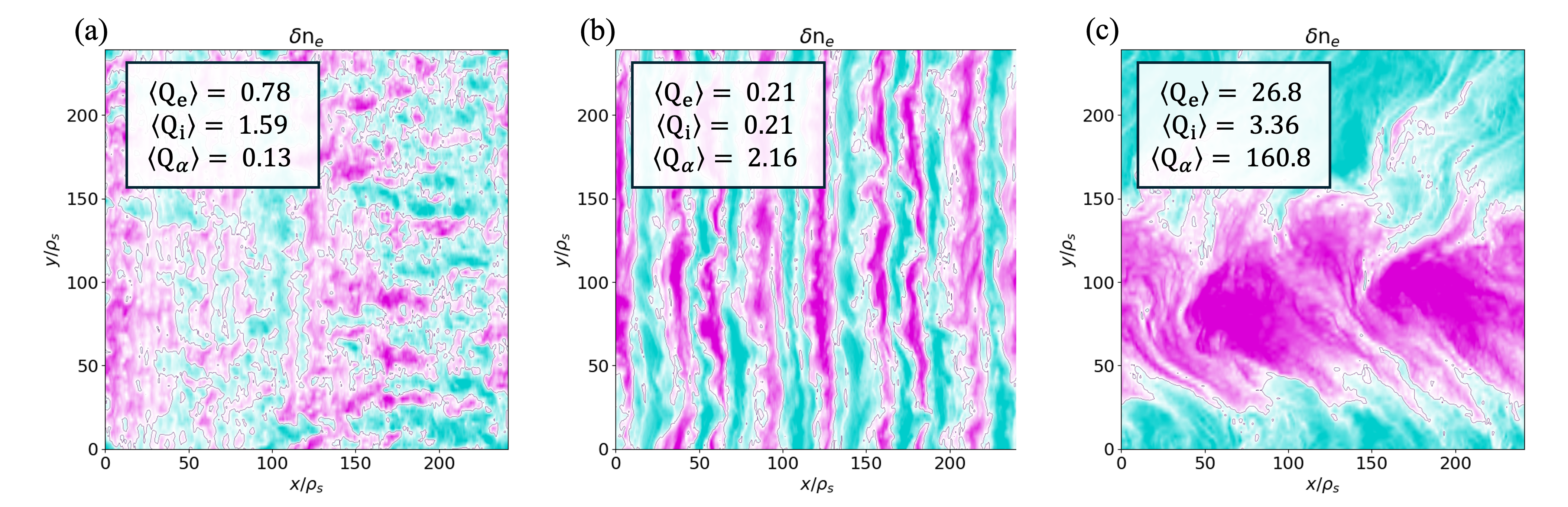}
    \caption{Real-space density fluctuations for nonlinear \CGYRO simulations at $r/a=0.35$ with $a/L_{T_i}$ +20\% relative to \MAESTRO profiles. (a) Nominal alpha density and normalized density gradient. (b) 50\% increased alpha density, showing destabilized fast ion mode and strongly suppressed turbulence. (c) 50\% increased alpha density gradient, showing a strongly unstable box-scale fast ion mode driving high alpha heat flux.}
    \label{fig:alphascansdensityfluct}
\end{figure*}

The simulation with artificially increased alpha density is effectively below the ITG critical gradient due to the increased linear drive for the fast ion mode, as well as due to direct stabilization due to the increase in $\beta$ and $\beta^*$. For the +50\% increased $n_\alpha$ case the change in $n_{DT}/n_e$ is $\lt0.4\%$, largely ruling out dilution effects as a cause of the reduced heat fluxes. In contrast, there is no significant change in heat fluxes when the fast alpha density gradient is increased until a critical threshold is reached and the instability becomes the dominant box-size effect at $a/L_{n_\alpha}+50\%$, shown in Figure \ref{fig:alphascansdensityfluct} (c). At this dramatically increased alpha density gradient, it is unlikely that the box-size effects are accurately captured by local simulations. However, it should be emphasized that such high alpha density gradients are likely not physical due to the high alpha heat and particle fluxes that would be driven by this very unstable fast ion mode. In the context of an FPP, as the alpha density gradient itself cannot be easily changed via external actuators, it is promising that increased heating and higher fusion power may be able to increase the degree of turbulence stabilization simply due to the increased alpha density, leading to a beneficial cycle. It is important to again emphasize here that we have not considered here the anomalous transport of alpha particles by fast-ion instabilities in determining fast alpha profiles. Whether the fast ion stabilization of turbulence persists when alpha pressure profiles are self-consistent with transport by TAEs and other fast ion modes is an important direction that will be addressed in future work.

\subsection{Scans in plasma Beta}

\begin{figure}
    \centering
    \includegraphics[width=0.9\linewidth]{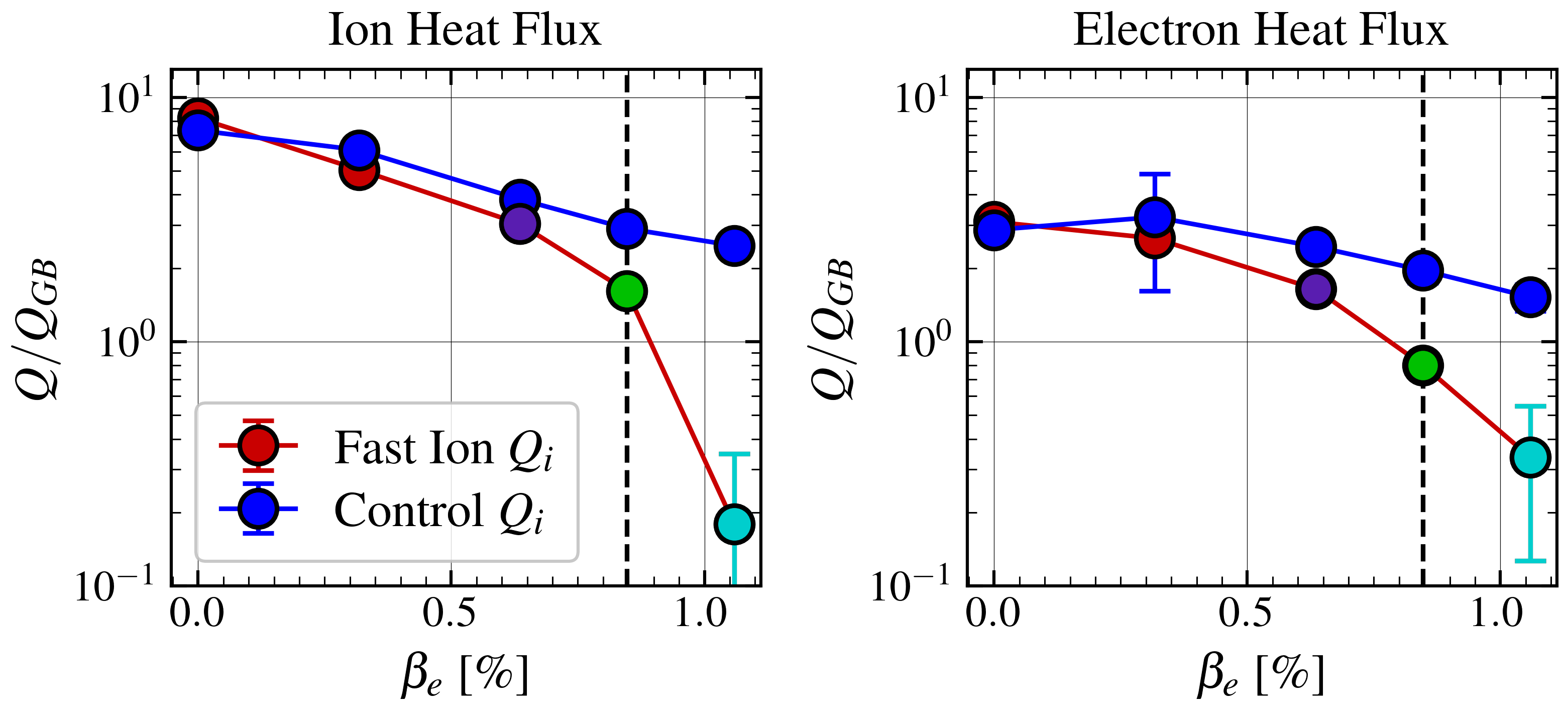}
    \caption{Nonlinear ion and electron heat flux for different values of $\beta_e$ for the simulation at $r/a=0.35$ with $a/L_{T_i}$ +20\% relative to \MAESTRO profiles. Heat fluxes decrease with $\beta_e$ for the control case due to electromagnetic stabilization of the ITG mode. For the fast ion simulations, the ITG stabilization is greater then the control cases for all simulations with $\beta_e>0$, and above the nominal value of $\beta_e$ (denoted with a dashed vertical line) exhibits a strong decrease in heat flux due to a strongly destabilized fast ion mode. Colors indicate relationship to the heat flux spectra shown in Figure \ref{fig:betaspectra}.}
    \label{fig:beta}
\end{figure}

\begin{figure*}
    \centering
    \includegraphics[width=\linewidth]{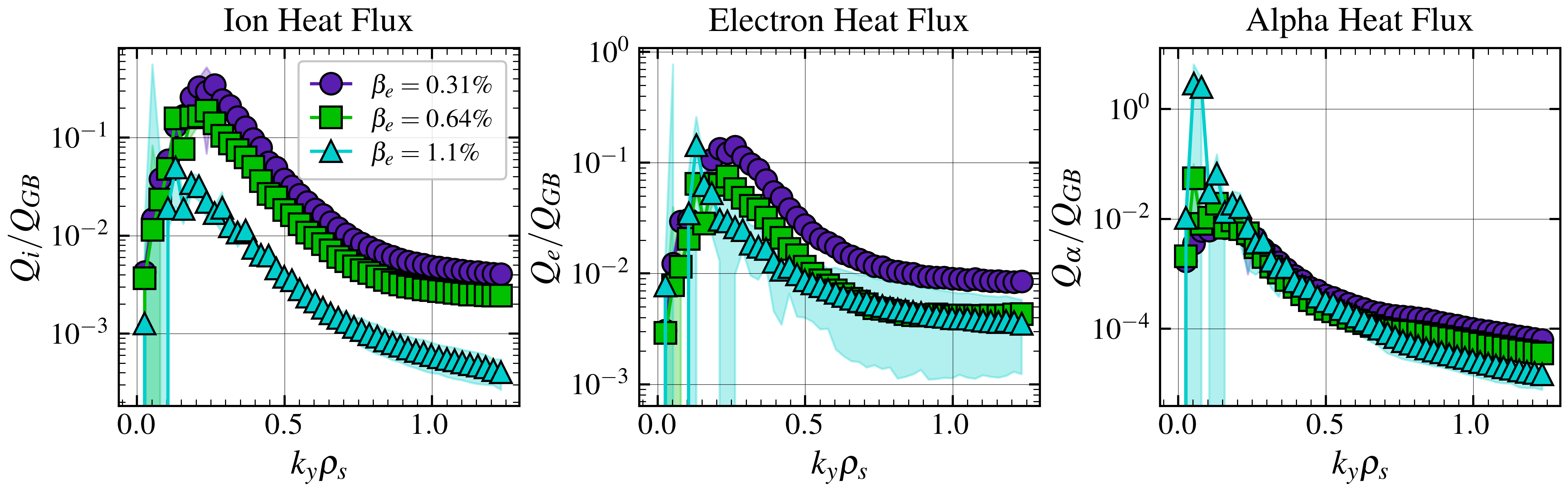}
    \caption{Spectrum of nonlinear fluxes as a function of binormal wavenumber $k_y\rho_s$ for cases at 75\%, 100\%, and 125\% the nominal value of $\beta_e$. As heat fluxes decrease with increasing $\beta_e$, the corresponding alpha particle heat flux increased and becomes peaked between $k_y\rho_s=0.05-0.01$, suggesting that destabilization of beta-driven fast ion modes is responsible for the decreased heat fluxes.}
    \label{fig:betaspectra}
\end{figure*}

As part of identifying the specific mechanism of turbulence stabilization, a scan of plasma $\beta_e$ was performed with nonlinear \CGYRO simulations. The normalized pressure gradient $\beta^*$ is changed self-consistently with the increase in $\beta_e$. The results, presented in Figure \ref{fig:beta}, demonstrate that the observed turbulence stabilization disappears in the electrostatic limit. It is somewhat surprising that the heat flux is such a sensitive function of $\beta_e$ despite the turbulent heat transport being primarily electrostatic (at $r/a=0.35$, $\rvert Q_{EM}/Q\rvert \lt5\%$). As $\beta_e$ is increased, the typical finite-beta stabilization of the ITG instability is observed, and the heat fluxes for both fast ion and control simulations decrease at similar rates. At values of $\beta_e$ above 1\%, the heat fluxes strongly diverge, likely due to the sub-dominant fast ion mode becoming unstable, as was observed at the nominal alpha pressure gradient. Until this critical point is reached, the heat fluxes diverge slowly for the fast ion vs. control simulations. This physical picture is supported by the alpha particle heat flux spectra, which shows significant peaking at the most unstable wavenumber for the increased $\beta_e$ case (Figure \ref{fig:betaspectra}). These results underscore the importance of including full electromagnetic effects, even in fusion power plant scenarios where the dominant instability is electrostatic ITG.


\subsection{Radial Scans}

\begin{figure*}
    \centering
    \includegraphics[width=\linewidth]{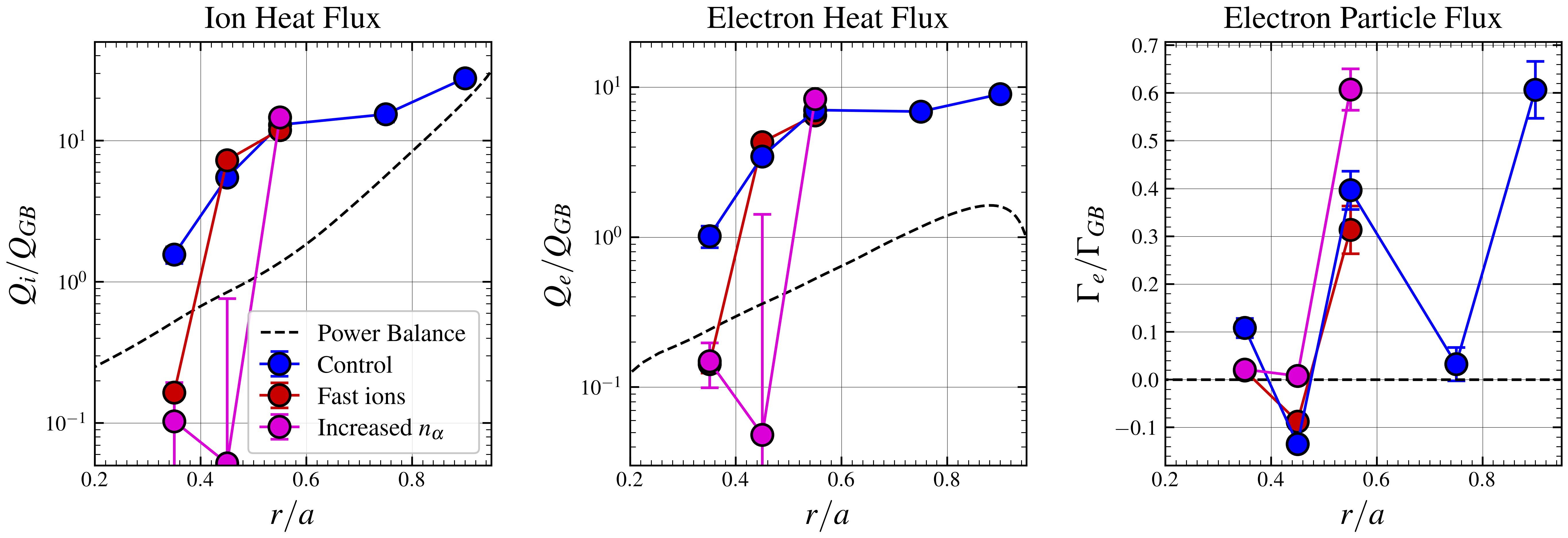}
    \caption{Comparison on nonlinear heat and particle turbulent fluxes without fast alphas, with fast alpha density predicted by \NUBEAM, and with fast alpha density increased by 50\%. Black dashed line denotes the target power balance flux for profiles produced from \MAESTRO integrated modeling. Compared to the nominal case, the simulations with increased alpha density show reduced heat fluxes and particle fluxes out to to r/a=0.45, but moderately higher heat and particle fluxes at r/a=0.55.}
    \label{fig:radial}
\end{figure*}

In order to determine the potential impact of the observed stabilization effects on global performance, it is necessary to know the radial extent of fast alpha particles' impact on turbulence. Three sets of radial location scans were performed: One control scan, one scan with the nominal fast alpha density and effective temperature taken from \NUBEAM, and one with the fast alpha density increased by 50\%. The increased alpha density is intended to emulate the fast alpha profile present in a full-current ARC discharge, based on results from \cite{howardPerformanceTransportARC2026}, and is meant to suggest whether the stabilization effects would extend further out at higher fusion power. The following radial locations were simulated with fast alpha particles: $r/a$$=[0.35,0.45,0.55]$. These choices are motivated from a database study of Alcator C-mod published in \cite{rodriguez-fernandeEnablingFirstprinciplesPredictions2025}, where it was found that a piecewise linear parametrization of the temperature and density gradients at $r/a$ $=[0.35,0.55,0.75,0.875,0.9]$ was sufficient to reproduce experimental fusion power within uncertainty. Here, we add a simulation at $r/a=0.45$, because of the possibility that alpha-particle induced stabilization will lead to more structure in the temperature and density gradient profiles. Nonlinear saturated heat fluxes from \CGYRO as a function of radius are plotted in Figure \ref{fig:radial}. Consistent with expectations from linear modeling, the observed turbulence stabilization at $r/a=0.35$ is not present at any outer radial locations. In fact, turbulent heat fluxes at $r/a=0.45$ are slightly higher compared to control simulations, though the effect is relatively small.

\begin{figure}
    \centering
    \includegraphics[width=\linewidth]{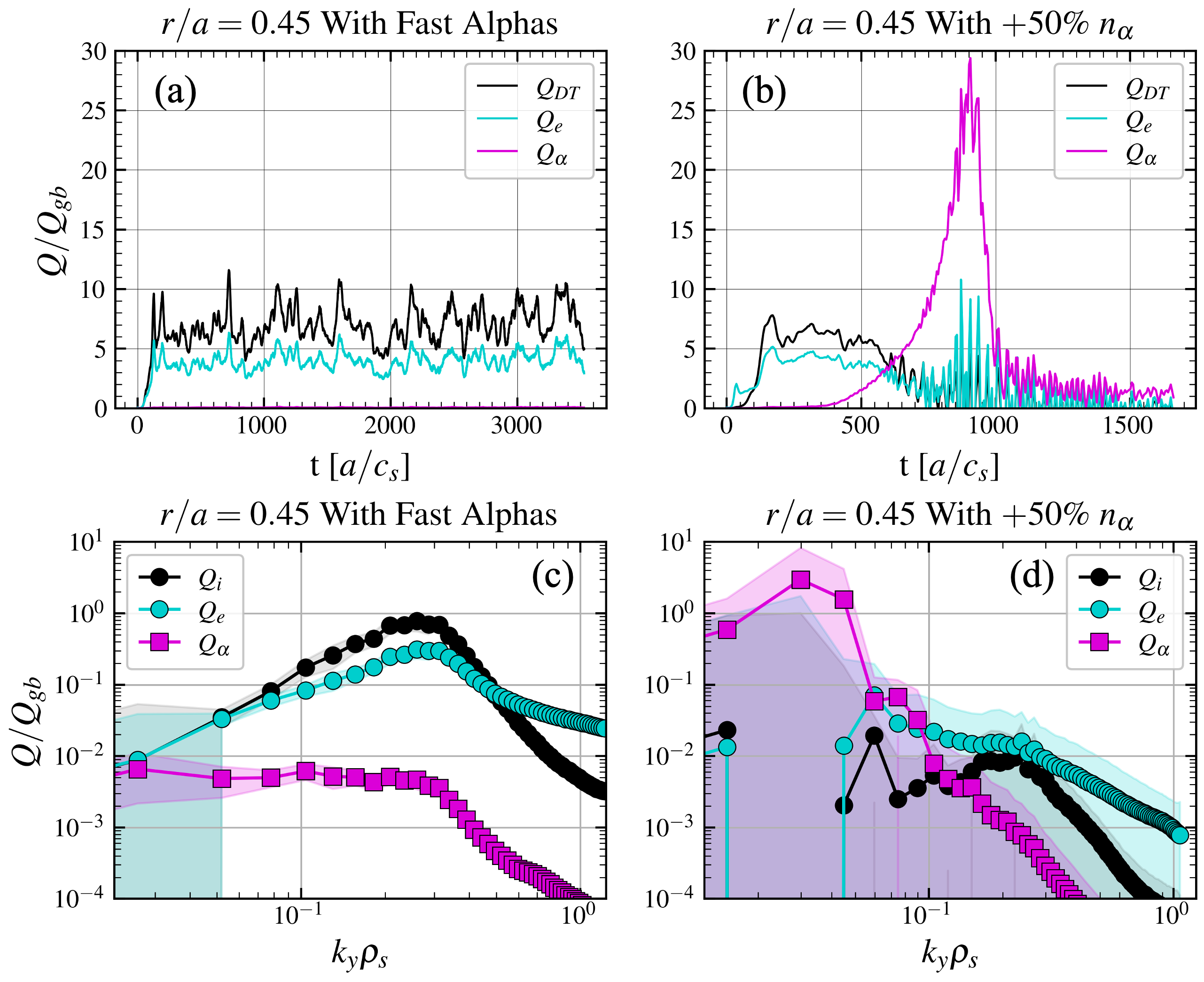}
    \caption{Nonlinear heat flux and heat flux spectra for $r/a=0.45$, for both the nominal fast alpha density from \NUBEAM (a, c) and 50\% increased alpha particle density (b,d). The growth of a fast alpha-destabilized mode is observed to strongly reduce the predominantly ITG turbulence in the increased fast alpha case, corresponding to greatly increased alpha heat fluxes, peaked at $k_y\rho_s=0.032$ ($n\sim13$). }
    \label{fig:increasedfast045}
\end{figure}

In the case of increased alpha density, however, stronger turbulence stabilization was observed, extending to $r/a=0.45$. Consistent with the results in Section \ref{subsec:alphascans}, the increased alpha density produces a strong reduction in turbulence without a change in the alpha pressure gradient (since the alpha density is simply multiplied by a scalar value, the normalized gradients for both nominal and increased fast ion cases are the same). However, heat and particle fluxes were slightly larger at $r/a=0.55$ with increased $n_\alpha$, showing that increased fast ion mode drive does not universally suppress heat fluxes. It should be mentioned that a minimum $k_y\rho_s=0.026$ was insufficient to properly capture low-n fast-ion instabilities for $r/a=0.45$, and additional convergence testing was done to determine a satisfactory resolution for this scan. The minimum and maximum wavenumbers simulated for these nonlinear simulations were $k_y\rho_s=[0.015,1.061]$ for $r/a=0.45$ (minimum $n\sim6$) and $k_y\rho_s=[0.015,1.153]$ (minimum $n\sim7$) for $r/a=0.55$. An additional simulation was completed with a minimum $n=5$ ($k_y\rho_{s,\text{min}}=0.0125$), which did not lead to a significant change in saturated heat fluxes. A comparison of the heat fluxes with nominal and increased alpha density is presented in Figure \ref{fig:increasedfast045}. The physical picture is similar to the above simulations at $r/a=0.35$, in which the growth of a $140$ kHz fast-ion driven mode at low wavenumber corresponds to a large reduction in the heat flux to nearly zero. There is again a strong suppression of the main ion heat flux at $k_y\rho_s=0.032$ ($n\sim13$), where fast alpha particle flux is strongest. The fact that the stabilization effect is observed at larger radii with an alpha particle density representative of the full-current ARC V3A design is a positive signal that these effects may play a significant role in the performance of burning plasmas not captured by traditional modeling.

\section{Discussion}
\label{sec:Discussion}

While the simulations presented here represent some of the highest-fidelity, computationally demanding local gyrokinetic simulations of a burning plasma performed to date, there are several key physics gaps which must be addressed before this work is able to confidently predict the impact of fast alpha particles on device performance. We again emphasize that a standalone study with nonlinear gyrokinetics is necessary, but not sufficient to understand the impact of alpha particle stabilization effects on overall fusion performance. Future work will focus on flux-matched predictions of ARC fusion performance with \PORTALS-\CGYRO, including for the first time the beneficial effect of turbulence stabilization as well as the detrimental effect of alpha transport by fast ion modes, such as those observed in this work. 

It will be crucial for any self-consistent FPP performance prediction to incorporate not only the potentially beneficial effects of turbulence stabilization, but also the potentially detrimental effects of anomalous fast particle transport. Recent analyses of alpha particle stabilization using global gyrokinetics have identified high heat and particle energetic particle fluxes associated with TAE activity, \cite{disienaUnderstandingTurbulenceSuppression2025, disienaRoleAlphaParticles2025} but no work yet to our knowledge has self-consistently captured the effect of this fast particle profile relaxation on fusion performance. Several mature workflows exist for predicting alpha particle redistribution due to saturated AE activity, including the Resonance-Broadened Quasilinear (RBQ) model, \cite{gorelenkovVerificationApplicationResonance2019} which has been coupled to TRANSP to predict alpha profile relaxation in ITER. \cite{gorelenkovFastIonRelaxation2024} While alpha losses are not expected to pose a risk for ARC performance, the redistribution of fast particles could have major impacts on fast ion instability drive and subsequent effects on turbulence and thermal profiles.

The simulations presented here take advantage of the local (flux-tube) gyrokinetic approximation for greatly increased computational efficiency and eliminating the need for \textit{ad hoc} boundary conditions. We argue that the local approximation is particularly suited to simulating high-energy alphas, as enforcing a periodic boundary condition ensures that fast alphas with a large gyroradius see constant background turbulence and do not have the potential to interact with the radial buffer zones needed in global simulation. In addition, the very slow-growing fast ion modes observed in these simulations can be simulated with local \CGYRO for several thousands of $a/c_s$ usually in less than 1000 GPU-hours (typically 1-2 days of wall clock time), a feat which would be all but impossible with global simulations of similar resolution. A major limitation of global simulations in the context of flux-matched profile predictions is that great care must be taken to ensure the source terms for heat and particles are physical and representative of the overall power balance, and that the radial domain can rarely be made large enough to cover the entire plasma radius. In contrast, local simulations do not require source terms and can be performed across the full radial domain. 

In the cases presented here, care was taken with simulation domains to capture all large-scale radial potential fluctuations, avoiding potential box-size effects. The high field and correspondingly small alpha particle gyroradius of future high-field devices also benefit the accuracy of the local approximation. For these profiles, the alpha particle normalized gyroradius $\rho^*_\alpha=\sqrt{\langle T_{\alpha,\text{eff}}\rangle/m_\alpha}/\Omega_\alpha=7.36\times 10^{-3}$, which is comparable to the average main ion $\rho^*$ in existing confinement databases where the local approach is well-validated. \cite{verdoolaegeUpdatedITPAGlobal2021} This lends confidence to our use of local simulations, which are formally equivalent to the global approach in the limit of vanishing $\rho^*$ \cite{candyLocalLimitGlobal2004}. A rigorous study of the agreement between local and global simulations with fast ions in the $\rho^*\to0$ limit has not yet been conducted to our knowledge, and is urgently needed to address the observed discrepancies in the literature. As both local and global gyrokinetic simulations have been successfully validated against present-day fusion devices, it will be important in future work to additionally validate the local gyrokinetic approach in capturing fast-ion effects.

It has been argued that local gyrokinetic simulations, because they do not capture the finite radial extent of TAE and other fast-ion driven modes, underpredict the effects of fast ion suppression of turbulence, especially when the linear fast ion instability is strongly driven \cite{disienaHowAccurateAre2023}. In an off-axis ICRH heating SPARC scenario in which significant fast-ion stabilization was present \cite{disienaPredictionsImprovedConfinement2023}, local simulations agreed well with global simulations, except at locations with a large fast ion temperature gradient and resonant interaction between the ICRH-accelerated $^3\text{He}$ minority and the background plasma. At the resonance locations, local simulations predicted significantly higher turbulent fluxes than global simulations, though for a predictive simulation a comparison to experimental ground truth is, of course, not available. However, these simulations in SPARC did not account for the redistribution of fast particles by the predicted fast ion instabilities, and it is therefore likely that the linear instability is not steady-state and would redistribute fast particles, relaxing the fast ion gradient. In the steady state, by necessity the fast ion-destabilized modes are marginally stable due to the fast ion profile having fully relaxed. In this state, local simulations are likely to be an effective approximation of the global heat flux. This highlights the importance of truly flux-matched, multi-channel profile predictions in both the thermal and fast particles. To our knowledge this is as yet an unexplored topic due to the large computational cost and multi-scale, multi-physics modeling needed. 

It should be mentioned that there exists a formalism for introducing second-order profile variation effects in local simulations with \CGYRO \cite{candyParadigmGlobalGyrokinetic2025, candySpectralTreatmentGyrokinetic2020}, which is formally equivalent to increasing the accuracy of the local approximation by an additional order in $\rho^*$ \cite{parraEquivalenceTwoDifferent2015}. While global profile shear effects were found to introduce only perturbatively small corrections to local simulations, it is unknown whether the these effects will still be negligible in the presence of fast alpha particles. This global-spectral formulation of local gyrokinetics could be used in future work to directly investigate the discrepancies arising between local and global simulations, without the confounding effects of differing numerical representations of heat and particle sources and \textit{ad hoc} boundary conditions. 

\section{Conclusions}

In the preceding sections, we have presented evidence that fast alpha-driven modes in the inner core of a burning plasma tokamak can strongly interact with and stabilize ion-scale turbulence by increasing the zonal flow shearing rate. The main results are summarized for convenience:

\begin{itemize}
    \item A significant reduction in ion-scale turbulent fluxes was observed in a reduced current ARC scenario in the presence of fast alpha particles.
    \item The observed stabilization can be effectively characterized as an upshift in the critical ITG gradient due to the fast ion mode-generated zonal flows.
    \item Turbulence stabilization was only observed in the presence of an unstable fast ion instability; direct stabilization effects were shown to be comparatively minor.
    \item The observed turbulence suppression scales beneficially with increased $\beta_e$ and $n_\alpha$.
\end{itemize}

This work motivates the use of nonlinear gyrokinetics, including fast alpha particles and self-consistent alpha transport, in flux-matched profile predictions of fusion power plants.

\section*{Acknowledgments}

The authors would like to extend thanks to G.J. Wilkie and N.N. Gorelenkov for useful discussions during the research process. This work was supported by Commonwealth Fusion Systems under RPP020 and by the MIT Office of Research Computing and Data Seed Fund program. The work of J.R.R. was partially supported by the UK's Engineering and Physical Sciences Research Council (EPSRC) [EP/W026341/1]. In addition we would like to gratefully acknowledge the use of the SDSC COSMOS system, an NSF user facility supported under NSF OAC-2404323. This research also made use of computing resources at the National Energy Research Scientific Computing Center (NERSC), a Department of Energy User Facility under NERSC award FES-ERCAP m4509 and m3195. \CGYRO is developed and maintained by the U.S. Department of Energy, Office of Science, Office of Fusion Energy Sciences under award DE-SC0024425.

\section*{Data Availability Statement}

The data that support the findings of this study are available upon reasonable request from the authors.

\bibliography{references.bib}

\end{document}